\documentclass[preprintnumbers,notitlepage,a4paper,aps,prd,onecolumn,superscriptaddress,nofootinbib,groupedaddress]{revtex4}

\usepackage{amsmath}
\usepackage{amsfonts,color}
\usepackage{amsthm}
\usepackage{pdfpages}
\usepackage{empheq}
\usepackage{amssymb,float}
\usepackage{booktabs,multirow}
\usepackage{titlesec}
\usepackage{hyphenat}
\usepackage{ragged2e}
\usepackage{amsmath}
\usepackage{accents}
\usepackage[utf8]{inputenc}
\usepackage{color}
\usepackage{hyperref}
\usepackage{enumitem}
\usepackage{tikz}
\usetikzlibrary{shapes.geometric}
\usetikzlibrary{arrows.meta,arrows}
\usepackage{mathrsfs}
\usepackage{ulem}

\titleformat{\paragraph}
{\normalfont\normalsize\bfseries}{\theparagraph}{1em}{}
\titlespacing*{\paragraph}
{0pt}{3.25ex plus 1ex minus .2ex}{1.5ex plus .2ex}

\allowdisplaybreaks
\hypersetup{
    colorlinks=true,
    linkcolor=blue,
    filecolor=magenta,      
    citecolor=blue
}

\begin{document}

\preprint{KEK-TH-2736, KEK-Cosmo-0385}

\title{Revisiting Coincident GR in Internal STEGR Formulation}

\author{Kyosuke Tomonari}
\email[]{ktomonari.phys@gmail.com}
\affiliation{Interfaculty Initiative in Information Studies, Graduate School of Interdisciplinary Information Studies, The University of Tokyo. 7-3-1 Hongo, Bunkyo-ku, Tokyo 113-0033, Japan}

\author{Taishi~Katsuragawa}
\email[
(Corresponding author) 
]{taishi@ccnu.edu.cn}
\affiliation{Institute of Astrophysics, Central China Normal University, Wuhan 430079, China}

\author{Shin'ichi Nojiri}
\email[]{nojiri@nagoya-u.jp}
\affiliation{
KEK Theory Center, Institute of Particle and Nuclear Studies, 
High Energy Accelerator Research Organization (KEK), Oho 1-1, Tsukuba, Ibaraki 305-0801, Japan
}
\affiliation{
Kobayashi-Maskawa Institute for the Origin of Particles and the Universe, 
Nagoya University, Nagoya 464-8602, Japan
}

\begin{abstract}
We revisit Coincident General Relativity (CGR) in the gauge approach to gravity based on Symmetric Teleparallel Equivalent to General Relativity (STEGR) in the \textit{internal-space formulation}, which one of the authors recently proposed in Ref.~[J. Math. Phys. 66 (2025) 5, 052505]. 
First, we review the standard formulation of STEGR theories in the Palatini approach to gravity, in which formulation we impose the teleparallel and torsion-free conditions by using Lagrange multipliers. 
Second, we introduce the STEGR theories in the gauge approach to gravity, which is formulated in the internal space, and derive its field equations. 
We briefly discuss whether the Ostrogradski ghost instability exists and find that the theory may require a degenerate condition to be imposed. 
Finally, assuming the coincident gauge, we derive CGR in both terms of the action integral and the field equation. 
Discussing the possible kinematics of the STEGR theory, we formulate a motion of a test scalar particle in the spacetime with non-metricity.
\end{abstract}

\maketitle

\section{\label{01}Introduction}

Symmetric Teleparallel Equivalent to General Relativity (STEGR) describes gravity using non-metricity in flat and torsion-free spacetime~\cite{Nester:1998mp, Bahamonde:2021gfp} and stands in Metric-Affine gauge theories of Gravity (MAG)~\cite{Hehl:1994ue}. 
The action of STEGR is equivalent to GR up to the total derivative, and thus, they are equivalent at the level of the field equation.
Providing an alternative geometric description of gravity to GR, STEGR may offer new perspectives on fundamental issues such as the quantum nature of gravity~\cite{Kiefer:2004xyv} and spacetime singularities~\cite{Hawking1973, Wald:1984rg}, which remain unresolved within GR.
However, recent progress has revealed several issues with the theoretical formulation of STEGR in the gauge approach to gravity, particularly in the coincident gauge, known as Coincident GR (CGR)~\cite{BeltranJimenez:2017tkd, Heisenberg:2023lru}.
The coincident gauge imposes a vanishing affine connection, so the STEGR is written solely in terms of the metric. 
Although the coincident gauge can simplify the description of STEGR, it also causes several issues in cosmological applications~\cite{Gomes:2023hyk, Gomes:2023tur, Heisenberg:2023wgk}

Among the issues discussed in recent studies of STEGR, an important one concerns its formulation within the gauge approach to gravity. 
In this work, we focus on this aspect.
STEGR is characterized by the non-metricity tensor $Q_{\rho \mu \nu}$, instead of the curvature tensor $R^{\lambda}_{\ \rho \mu \nu}$ in GR.
In the conventional formulation of the STEGR, we face a simple but important result regarding a gauge fixing:
When we employ the Weitzenb\"{o}ck connection in which the connection 1-form vanishes (so-called Weitzenb\"{o}ck gauge~\cite{Weitzenboh1923}), both curvature and non-metricity automatically vanish, and only torsion remains~\cite{Obukhov:2002tm, Ferraro:2016wht, Blagojevic:2023fys}. 
Indeed, the non-metricity tensor in the Weitzenb\"{o}ck gauge always vanishes.
Although the Weitzenb\"{o}ck connection is convenient to establish Teleparallel Equivalent to GR (TEGR)~\cite{Einstein1928},
we cannot straightforwardly use the Weitzenb\"{o}ck connection in the STEGR.
This gauge-fixing issue indicates that it is necessary to explore a formulation of STEGR that remains well defined even in the Weitzenböck gauge.

Ref.~\cite{Tomonari:2024vij} proposed three new formulations of STEGR, denoted by Formalisms 1, 2, and 3, using the idea of an internal space (we will briefly review it in Sec.~\ref{addedSec}).
In the gauge approach to gravity, 
the internal-space metric is a gauge field, in addition to the frame field and the connection 1-form, allowing us to introduce a non-vanishing non-metricity even in the Weitzenb\"{o}ck gauge. In this approach, the torsion and non-metricity do not vanish in general, but only the latter must remain in STEGR. 
One can realize it by imposing the torsion-free property as a constraint on its Lagrangian. 
Formalisms 1 and 2 take into account this constraint as an outer (using Lagrange multipliers) and an inner (imposed by hand) primary constraint, respectively. 
Formalism 3 realizes this constraint differently. 
Decomposing the frame field into St\"{u}ckelberg fields, one finds that the torsion-free property is automatically satisfied, and the St\"{u}ckelberg fields play the role of a gauge field rather than the frame field. 

In this work, we shed light on the theoretical aspects of STEGR and examine STEGR in the gauge approach to gravity and its physical implications. 
The construction of the paper is given as follows: 
In Sec.~\ref{addedSec}, we introduce the basics of the gauge approach to gravity.
In Sec.~\ref{02}, we review the STEGR theories focusing on their formulations. 
In Sec.~\ref{03}, we investigate STEGR theories in the gauge approach based on the internal STEGR in Formalism 3 proposed recently in Ref.~\cite{Tomonari:2024vij}. 
We will derive its field equations and correctly reformulate CGR.
The internal-space formulation clarifies the meaning of the coincident gauge as the gauge approach to gravity.
In Sec.~\ref{04}, we discuss the kinematics of STEGR theories. 
In particular, we formulate a method for deriving the trajectory equation and discuss a spacetime-dependent mass in terms of proper time. 
This suggests that a particular coordinate system may exist in which the magnitude of the mass remains constant when expressed with respect to a non-proper time.
Finally, in Sec.~\ref{05}, we summarize this work and provide future directions. 


\section{Rudiments for Gauge Approach to Gravity}
\label{addedSec}
In our formulation, we have two geometries: the spacetime $\mathcal{M}$ and the internal space $\mathcal{V}$. 
The internal space, $\mathcal{V}$, is the the total space of a vector bundle $(\mathcal{V},\mathcal{M},\iota)$, where $\iota$ is the projection map from $\mathcal{V}$ to $\mathcal{M}$.
The dimension of spacetime $\mathcal{M}$ is $n+1$. 
These geometries are related by a bundle morphism called the frame field $e:\mathcal{V} \rightarrow T\mathcal{M}$, where $T\mathcal{M}$ is the tangent bundle of spacetime. 
In this work, we assume that $e$ is locally isomorphic.
In this section, we briefly review and define geometric quantities for studying and formulating the STEGR in the gauge approach to gravity.\footnote{
For instance, see Refs.~\cite{Tomonari:2023ars, Tomonari:2024vij} .
}

First, we summarize basic geometric ingredients on the spacetime, which are defined by using the tangent bundle of spacetime $(T\mathcal{M},\mathcal{M},\rho)$, where $\rho$ is the projection map from $T\mathcal{M}$ to $\mathcal{M}$.
In the spacetime, we use an affine connection, $\Gamma^{\rho}_{\mu \nu}$, defined as
\begin{align}
\begin{split}
    \nabla_{\mu}\partial_{\nu} 
    := 
    \Gamma^{\rho}_{\mu \nu} \partial_{\rho}
\end{split}
\end{align}
and introduce the covariant derivative $\nabla_{\mu}$ on $(T\mathcal{M},\mathcal{M},\rho)$.
The affine connection is decomposed into
\begin{align}
    \Gamma^{\rho}_{\mu\nu} 
    = 
    \overset{\circ}{\Gamma}{}^{\rho}_{\mu\nu} + N^{\rho}{}_{\mu \nu}
    \,,
\label{decomposition of affine connection}
\end{align}
where $\overset{\circ}{\Gamma}{}^{\rho}_{\mu\nu}$ is the Levi-Civita connection.
In the following, we use a circle symbol $\overset{\circ}{}$ to denote quantities with respect to the Levi-Civita connection. 
$N^{\rho}{}_{\mu \nu}$ is called the distortion tensor defined as
\begin{align}
\begin{split}
    &N^{\rho}{}_{\mu \nu} := K^{\rho}_{\mu\nu} + L^{\rho}_{\mu\nu} \,,\\
    &K^{\rho}{}_{\mu\nu} := \frac{1}{2}T^{\rho}{}_{\mu\nu} - T_{(\mu\nu)}{}^{\rho} \,, \\
    &L^{\rho}{}_{\mu\nu} := \frac{1}{2}Q^{\rho}{}_{\mu\nu} - Q_{(\mu\nu)}{}^{\rho} \,.
\end{split}
\label{distorsion, contorsion, and disformation}
\end{align}
We call $K^{\rho}{}_{\mu\nu}$ and $L^{\rho}{}_{\mu\nu}$ the contorsion and disformation tensor, respectively.
$T^{\rho}{}_{\mu \nu}$ and $Q_{\rho \mu \nu}$ are the torsion and non-metricity tensors  defined as
\begin{align}
\begin{split}
    &T^{\rho}{}_{\mu \nu} = 2 \Gamma^{\rho}_{[\mu \nu]} \,,\\
    &Q_{\rho \mu \nu} = \nabla_{\rho}g_{\mu\nu} \, ,
\end{split}
\label{torsion and non-metricity}
\end{align}
where $g_{\mu \nu}$ is the spacetime metric and we used the anti-symmetrization $\mathscr{O}_{[XY]} = (\mathscr{O}_{XY} - \mathscr{O}_{YX})/2$.
The non-metricity plays the central role in this paper.
Relating to the affine connection, we have the generic curvature tensor defined as
\begin{align}
\begin{split}
     R^{\lambda}{}_{\rho \mu \nu} 
     &:= 
     2 \partial_{[\mu}\Gamma^{\lambda}_{\nu] \rho} 
     + 2 \Gamma^{\lambda}_{[\mu|\kappa|}\Gamma^{\kappa}_{\nu]\rho} \\
     &= 
     \overset{\circ}{R}{}^{\lambda}{}_{\rho\mu\nu}
     + 2 \overset{\circ}{\nabla}_{[\mu}N^{\lambda}{}_{\nu]\rho}
     + 2 N^{\lambda}{}_{[\mu|\kappa|}N^{\kappa}{}_{\nu]\rho}
     \,,
\end{split}
\label{decomposition of curvature}
\end{align}
where we applied Eq.~\eqref{decomposition of affine connection} in the second line. 
Contracting this generic curvature tensor, we derive
\begin{align}
    R 
    = 
    \overset{\circ}{R} 
    + 2\overset{\circ}{\nabla}{}_{\mu}N^{[\mu|}{}_{\nu}{}^{|\nu]} 
    + 2N^{\mu}{}_{[\mu|\nu|}N^{\nu}{}_{\rho]}{}^{\rho} 
    \,.
\label{generic scalar curvature}
\end{align}
In this geometry, the affine connection is a gauge field, which transforms as follows:
\begin{align}
    \Gamma'^{\rho}_{\mu\nu} 
    = 
    \frac{\partial x'^{\rho}}{\partial x^{\alpha}}
    \frac{\partial x^{\beta}}{\partial x'^{\mu}}
    \frac{\partial x^{\gamma}}{\partial x'^{\nu}}
    \Gamma^{\alpha}_{\beta\gamma} 
    + \frac{\partial x'^{\rho}}{\partial x^{\lambda}}
    \frac{\partial^{2} x^{\lambda}}{\partial x'^{\mu}\partial x'^{\nu}} \,
\label{Law of gauge transformation of affine connection}
\end{align}
for a coordinate transformation, $x^{\mu} \rightarrow x'^{\mu} = x'^{\mu}(x^{\mu})$.

Second, we summarize basic geometric ingredients on the internal space $(\mathcal{V},\mathcal{M},\iota)$. 
In an internal space, we use a connection 1-form, $\omega = \omega^{I}{}_{J \mu} \zeta_{I}\otimes\zeta^{J}\otimes dx^{\mu} \in \mathrm{Section}\left(\mathrm{End}(\mathcal{V})\otimes T^{*}\mathcal{M}\right)$, defined as
\begin{align}
    \mathcal{D}_{\mu}\zeta_{I} := \omega^{J}{}_{I \mu}\zeta_{J}
\label{connection 1-form}
\end{align}
and introduce the covariant derivative $\mathcal{D}_{\mu}$ on $(\mathcal{V},\mathcal{M},\iota)$.
$\zeta_{I}$ is a basis of a local trivial bundle $\mathcal{V}|_{U}\simeq U\times \mathbb{R}^{n+1}$ for an open set $U$ in $\mathcal{M}$. 
$\zeta^{I}$ is its dual basis. 
Here, $\mathrm{Section}(X)$ and $\mathrm{End}(Y)$ denote a space of sections of a bundle $X$ and of auto-morphisms of a bundle $Y$.
Using the connection 1-form, the curvature, torsion, and non-metricity are introduced as
\begin{align}
\begin{split}
    &\tilde{R} = d_\mathcal{D}\omega = d\omega + \omega\wedge \omega \,,\\
    &\tilde{T} = d_\mathcal{D}e^{-1}  = de^{-1} + \omega\wedge e^{-1} \,,\\
    &\tilde{Q} = d_\mathcal{D}\eta = d\eta + \omega\wedge\eta \,,
\end{split}
\label{curvature, torsion, and non-metricity in internal space}
\end{align}
where $g_{I J}$ is the internal-space metric, and $\eta = g_{I J} \zeta^{I}\otimes \zeta^{J}$. 
$d_\mathcal{D}$ is the covariant exterior derivative with respect to $\mathcal{D}$,
and $e^{-1}$ is the inverse frame field of $e$. 
In this geometry, the connection 1-form is a gauge field, which transforms as follows:
\begin{align}
    \omega'^{U_{j}} = (\tau^{-1})_{ji}d\tau_{ji} + (\tau^{-1})_{ji}\omega^{U_{i}}\tau_{ji}
\label{Law of guage transformation of connection 1-form}
\end{align}
for $s^{U_{i}} \rightarrow s'^{U_{j}} = \tau(p) s^{U_{i}}$, where $p\in U_{i}\cap U_{j}$, $s^{U_{i}}\in\mathrm{Section}(\mathcal{V}|_{U_{i}})$, and $U_{i}$ is an set of open cover of spacetime.
Then, we have $\mathcal{D}^{U_{j}}s' = \tau_{ji}(p) \mathcal{D}^{U_{i}}s$, where $G$ is the structure Lie group of $\mathcal{V}$ and $\tau_{ij} \in G$. 

For instance, if we consider the case of $G = SO(1,3)$, the connection 1-form turns out to be the spin connection, and the internal-space metric is restricted to the Minkowski one: $g_{I J} =  \eta_{I J} = \mathrm{diag}(-1,1,1,1)$. 
In symmetric teleparallel spacetimes, where curvature and torsion vanish, the internal gauge group $G$ remains as the group preserving the non-metricity structure. 
Under a gauge transformation $\tau \in G$, the internal metric transforms as $g_{IJ} \to \tau_I{}^K \tau_J{}^L g_{KL}$.
As we will introduce later, the St\"{u}ckelberg fields transform as $\xi^I \to (\tau^{-1})^I{}_J \xi^J$, and these transformations preserve the Weitzenb\"ock gauge locally.

Geometric quantities defined in the spacetime and internal space are related through the bundle morphism $e$ (and its inverse $e^{-1}$) as 
\begin{align}
\begin{split}
    &R = e(\tilde{R}) \,,\quad\mathrm{or}\quad R^{\sigma}{}_{\mu \nu \rho} = e_{I}{}^{\sigma} e^{J}{}_{\mu} \tilde{R}^{I}{}_{J \nu \rho} \,,\\
    &T = e(\tilde{T}) \,,\quad\mathrm{or}\quad T^{\rho}{}_{\mu \nu} = e_{I}{}^{\rho} \tilde{T}^{I}{}_{\mu \nu}\\
    &Q = e(\tilde{Q}) \,,\quad\mathrm{or}\quad Q_{\rho \mu \nu} = e^{I}{}_{\mu} e^{J}{}_{\nu} \tilde{Q}_{\rho I J}\,.
\end{split}
\label{relationships of geometric quantities between spacetime and internal space}
\end{align}
We emphasize that the vanishing of the spacetime curvature $R^\alpha{}_{\beta\mu\nu} = 0$
refers to the push-forward of the internal-space curvature via the frame fields.
Although the internal curvature $\tilde{R}^I{}_{J\mu\nu}$ may be nonzero, 
its push-forward $R^\alpha{}_{\beta\mu\nu} = e(\tilde{R})$ vanishes in the Weitzenböck gauge.
This explains why there is no contradiction between a non-vanishing internal curvature and a zero spacetime curvature.
The affine connection and the connection 1-form are related by
\begin{align}
     \Gamma^{\rho}_{\mu \nu} = e^{I}{}_{\mu}\partial_{\nu}e_{I}{}^{\rho} + \omega^{I}{}_{J \mu} e^{J}{}_{\nu} e_{I}{}^{\rho} \,.
\label{frame field postulate}
\end{align}
We remark that the above relation is equivalent to the imposition of 
\begin{align}
    \nabla v = \mathcal{D} \tilde{v} 
    \,.
\label{local condition of ACD}
\end{align}
$v = v^{\mu} \partial_{\mu}$ is the section of the tangent bundle of spacetime, and $\tilde{v} = (e^{-1})_{*}(v) = \tilde{v}^{I} \zeta_{I} = e^{I}{}_{\mu} v^{\mu} \zeta_{I}$ is the pushforward of $v$ by $e^{-1}$.
That is, we \textit{demand} this relation as a differential-geometric structure in our theory. 
This structure indicates that in the Palatini or gauge approach, we must take the configuration space as $Q_{\rm Palatini\,approach} = \left< g_{\mu \nu}\,,\Gamma^{\rho}_{\mu \nu}; \mathrm{variables\,related\,to\,constraints} \right>$ or $Q_{\rm Gauge\,approach} = \left< g_{I J}\,,\omega^{I}{}_{J\mu}\,,e^{I}{}_{\mu}; \mathrm{variables\,related\,to\,constraints} \right>$, respectively, as we mentioned in Sec.~\ref{01}. 

Under a coordinate transformation $x^\mu \to x'^\mu$, the affine connection transforms according to Eq.~\eqref{Law of gauge transformation of affine connection}, 
while the St\"{u}ckelberg fields transform as scalar functions: $\xi'^I(x') = \xi^I(x(x'))$. 
Therefore, $\Gamma$ in the new coordinates is generally nonzero. 
This does not contradict the construction in Eq.~\eqref{frame field postulate}, because the condition $\Gamma=0$ is gauge-dependent: it holds in the coincident gauge in a chosen coordinate system. 
The formalism distinguishes between coordinate and internal-gauge degrees of freedom, so there is no intrinsic inconsistency.

Finally, we introduce the Weitzenb\"{o}ck gauge~\cite{Weitzenboh1923, Blagojevic:2000pi, Ferraro:2016wht, Capozziello:2022zzh}, St\"{u}ckelberg field~\cite{Stueckelberg:1938a,Stueckelberg:1938b,Ruegg:2003ps}, and the coincident gauge~\cite{Adak:2008gd,Adak:2011ltj,DAmbrosio:2020nqu,BeltranJimenez:2022azb}.
We can restrict the covariant derivative of the internal space by imposing
\begin{align}
    \omega^{I}{}_{J} := 0 \, .
\label{Weitzenboeck gauge}
\end{align}
This is called the Weitzenb\"{o}ck gauge, corresponding to a gauge fixing of the internal connection.
When we set a coordinate system locally, we have
\begin{align}
    \omega^{I}{}_{J} = \omega^{I}{}_{J\mu}dx^{\mu} := 0 \,.
\label{Weitzenboeck gauge with coordinate system}
\end{align}
Since the one-forms $dx^{\mu}$ are linearly independent, we obtain $\omega^{I}{}_{J\mu} = 0$, which is the conventional form of the Weitzenb\"{o}ck gauge condition.
The Weitzenb\"ock gauge can always be imposed locally by a local gauge transformation in any open neighborhood. 
Globally, its existence depends on the trivializability of the vector bundle; in globally nontrivial bundles, it may hold only patchwise, requiring transition functions on overlaps.

Independently, we can also restrict the co-frame field by decomposing
\begin{align}
    e^{I}{}_{\mu} := \partial_{\mu}\xi^{I} \,,
\label{Stueckelberg field}
\end{align}
where $\xi^{I}$ for $I = 1,2,\cdots,n+1$ are a set of scalar functions, called St\"{u}ckelberg fields.
The introduction of the St\"{u}ckelberg fields reflects an underlying redundancy associated with internal coordinate freedom.
This redundancy can be fixed by identifying the internal coordinates with the spacetime coordinates up to a linear transformation, which defines the coincident gauge.
The spacetime metric $g_{\mu\nu}$ and internal metric $g_{IJ}$ are related by
\begin{align}
    g_{\mu\nu} 
    = 
    \partial_\mu \xi^I \, \partial_\nu \xi^J \, g_{IJ}
    \, .
\end{align}
This relation should be understood as a field reparametrization rather than a fundamental definition of the spacetime metric.

The restriction in Eq.~\eqref{Stueckelberg field} is equivalent to the introduction of the local geometric structure $\mathcal{V}|_{U} \simeq T\mathcal{M}|_{U} \simeq U\times\mathbb{R}^{n+1}$. 
We note that this decomposition breaks the \textit{local} symmetry because the gauge transformation acts on the index $I$ of $\xi^{I}$ instead of that of $e^{I}{}_{\mu}$. 
Inspired by $\mathcal{V}|_{U} \simeq T\mathcal{M}|_{U} \simeq U\times\mathbb{R}^{n+1}$, the St\"{u}ckelberg fields can be decomposed into 
\begin{align}
    \xi^{I} := A^{I}{}_{\mu} x^{\mu} + a^{I}  
\label{coincident gauge in generic form}
\end{align}
where $A^{I}{}_{\mu} \in GL(n+1;\mathbb{R})$, $x^{\mu}$ is the coordinate functions covering $U$, and $a^{I}$ is a constant spacetime vector. 
This is an affine transformation in $\mathbb{R}^{n+1}$. 
Geometrically, this decomposition can be interpreted as globalizing the internal space $\mathcal{V}|_{U}$ by $\mathcal{V}|_{\mathcal{M}}$.
In particular, the imposition of $A^{I}{}_{\mu} = \delta^{I}{}_{\mu}$ defines the coincident gauge, which was first proposed in the pioneering works~\cite{Adak:2008gd, Adak:2011ltj}.
Under this choice, the affine connection vanishes,
since Eq.~\eqref{frame field postulate} reduces to $\Gamma^\rho_{\mu\nu} = 0$
when $\omega^I{}_{J\mu}=0$ and $e^I{}_\mu = \partial_\mu \xi^I = \delta^I{}_\mu$.
Moreover, the dynamics of the spacetime metric decouple from the St\"{u}ckelberg fields, so that the spacetime metric obeys the same field equations as in GR.

We make the following remarks.
Residual internal gauge transformations that preserve the Weitzenb\"ock gauge correspond to constant elements of the original gauge group, $\tau = \Lambda \in G$, and act as affine transformations $\xi^I \to \Lambda^I{}_J \xi^J + a^I$, 
where $\Lambda^I{}_J$ and $a^I$ are constants, since spacetime-dependent transformations would reintroduce a non-vanishing connection through the inhomogeneous term in the gauge transformation law. 
These transformations do not affect the affine connection or the dynamics of the metric.

We comment on the interpretation and global realization of the coincident gauge in relation to the residual symmetry structure above.
The condition $\omega^I{}_{J\mu}=0$ can be imposed as a local statement independently of the global trivializability of the bundle, although its global realization depends on the global properties of the connection.
The coincident gauge $\xi^I = x^\mu \delta^I_\mu$ can then be introduced locally on top of the Weitzenb\"ock gauge. 
Its global implementation is not guaranteed in general and depends on the global structure of the fields and the underlying manifold. 
It should be emphasized that the coincident gauge is not a gauge in the conventional sense of an internal symmetry transformation; rather, it corresponds to a choice of coordinates together with a particular identification of the St\"{u}ckelberg fields with spacetime coordinates.


\subsection{Relations between Internal-Space and Spacetime}

By introducing the internal space in the gauge approach to gravity, a natural bi-metric structure arises. 
In particular, the relation between the frame field $e_I{}^\mu$ and the St\"{u}ckelberg fields $\xi_A$ generates a formal chain~\cite{Tomonari:2024vij}: 
\begin{align}
    e_{A}{}^{\mu} 
    = 
    \partial^{\mu}\xi_{A} 
    = 
    e_{I}{}^{\mu}\,e_{J}{}^{\nu}\,g^{IJ}\,\partial_{\nu}\xi_{A} 
    = 
    \partial^{\mu}\xi_{I}\,\partial^{\nu}\xi_{J}\,g^{IJ}\,\partial_{\nu}\xi_{A} 
    = g^{\mu\rho}\,g^{\nu\lambda}\,g^{IJ}\,\partial_{\rho}\xi_{I}\,\partial_{\lambda}\xi_{J}\,\partial_{\nu}\xi_{A} 
    = 
    \cdots
    \,,
\end{align}
indicating that $\partial^\mu$ must be interpreted via the spacetime metric, i.e., $\partial^\mu = g^{\mu\nu}\partial_\nu$.  
This implies that the spacetime metric $g_{\mu\nu}$ cannot be eliminated and effectively acts as a multiplier,  
while the internal-space metric $g_{IJ}$ serves as the dynamical variable describing gravity.  
We will further clarify the manifestation of this bi-metric structure in Sec.~\ref{03:01}.

The spacetime metric $g_{\mu\nu}$ and the internal metric $g_{IJ}$ are related through the frame field $e^I{}_\mu$, but not via a naive pullback $g_{\mu\nu} = e^{A}{}_{\mu} e^{B}{}_{\nu} g_{AB}$. 
Instead, the bundle morphism enforces a relation at the level of sections as in Eq.~\eqref{local condition of ACD}.
This ensures consistency between spacetime and internal-space covariant derivatives while keeping $g_{\mu\nu}$ independent as a multiplier-like field.

In the conventional formulation, we use the Palatini approach to formulate STEGR~\cite{BeltranJimenez:2019esp} and introduce the action integral of STEGR with two constraints, the teleparallel and torsion-free conditions, which we impose using Lagrange multipliers. 
As a result, this theory describes gravity in terms of the metric and the affine connection. 
No issues arise at this stage; however, we shall encounter a fault in the formulation when we establish CGR or work in the coincident gauge. 
Using the St\"{u}ckelberg fields to decompose a frame field, one finds that the torsion automatically vanishes~\cite{BeltranJimenez:2022azb}.
This is the same as in the internal-space formulation, and the non-metricity arises from the violation of the \textit{local} Lorentz symmetry. 
Imposing further a gauge condition that the St\"{u}ckelberg fields coincide with the coordinate functions of spacetime, we obtain CGR. 
However, this conventional approach formulates the CGR with \textit{misleadingly} applying the gauge-fixing method, which is valid only in the gauge approach. 

In this conventional formulation, the Minkowski metric is implicitly chosen as the internal-space metric, and the non-metricity is regarded as 
\begin{align}
    Q_{\rho \mu \nu} = \partial_{\rho}g_{\mu \nu} \,.
\label{conventional Q with C-gauge and Minkowski-gauge in wrong}
\end{align}
We emphasize that this is an inappropriate expression. 
In fact, we can further calculate as follows:
\begin{align}
\begin{split}
    Q_{\rho \mu \nu} 
    &= \partial_{\rho}g_{\mu \nu} \\
    &= \partial_{\mu}\xi^{I} \partial_{\nu}\xi^{J} \partial_{\rho}\eta_{I J} 
    + \partial_{\nu}\xi^{J} \partial_{\rho}\partial_{\mu}\xi^{I} \eta_{I J}
    + \partial_{\mu}\xi^{J} \partial_{\rho}\partial_{\nu}\xi^{I} \eta_{I J} \\
    &= \delta^{I}{}_{\mu} \delta^{J}{}_{\nu}\partial_{\rho}\eta_{I J} = 0 \,.
\end{split}
\label{conventional Q with C-gauge and Minkowski-gauge in correct}
\end{align}
In the final line, we applied the coincident gauge: $\xi^{I} = \delta^{I}{}_{\mu}x^{\mu}$. 
As will be discussed in detail in Sec.~\ref {03:02}, we can conclude a statement: To treat a theory as the gauge/Palatini approach in a \textit{well-posed} manner, we must formulate the theory in terms of internal space/spacetime variables only, respectively, except for the multipliers. 

Here, the word ``well-posed'' means that the configuration space of a given theory is composed of geometric quantities that relate to a common bundle and of variables that relate to constraints in the bundle.
The Palatini approach sets the configuration space as 
$Q_{\rm Palatini\,approach} = \left< g_{\mu \nu}\,,\Gamma^{\rho}_{\mu \nu}; \mathrm{variables\,related\,to\,constraints} \right>$.
The spacetime metric and the affine connection, respectively, are related only to a common bundle, and the tangent bundle of spacetime is $(TM,M,\rho)$. 
On the other hand, the gauge approach sets the configuration space as $Q_{\rm Gauge\,approach} = \left< g_{I J}\,,\omega^{I}_{J\mu}\,,e^{I}{}_{\mu}; \mathrm{variables\,related\,to\,constraints} \right>$.
The internal-space metric, the connection 1-form, and the frame field, respectively, are related only to a common bundle, and the internal bundle of spacetime is $(\mathcal{V},M,\iota)$. 

If we were confused about the above setup involving two bundles related by a bundle morphism, the independence of the configuration variables would not hold. 
In the conventional formulation of STEGR, we use the configuration space $Q_\mathrm{conventional\,STEGR} = \left<g_{\mu\nu},\xi^{I} \right>$, and this confusion leads us to the wrong expression of Eq.~\eqref{conventional Q with C-gauge and Minkowski-gauge in wrong} and prevents us from performing the canonical analysis of STEGR correctly. 
In Sec.~\ref{03:02}, based on Formalism 3 of the internal STEGR formulation, we will revisit CGR and verify that our theory correctly derives CGR from the point of view of the action integral and the field equation. 
We will also reveal that our theory may contain Ostrogradski's ghost degrees of freedom (DOF) with respect to the St\"{u}ckelberg fields.
Thus, to make our theory healthy, we must perform the canonical analysis of our theory to identify the degenerate condition, which removes this ghost DOF~\cite{Motohashi:2016ftl}. 

We summarize the current status of theoretical formulations of STEGR and CGR, comparing them with GR and TEGR, in Table~\ref{table: Theories and status in first group issue}. 
\begin{table}[ht!]
    \centering
    \renewcommand{\arraystretch}{1.5}
    \caption{
    Summary of the theoretical formulations considered in this work. 
    $g_{\mu\nu}$ and $\Gamma^{\rho}_{\mu\nu}$ are the spacetime metric and affine connection, respectively. 
    $g_{IJ}$, $e^{I}{}_{\mu}$, and $\xi^{I}$ denote the internal metric, co-frame field components, and St\"{u}ckelberg fields (four scalar fields), respectively. 
    The column “Gauge choice’’ indicates the gauge fixing applied to the standard covariant formulations. 
    `$\checkmark$' denotes a well-posed formulation, `$\times$' denotes the lack of a well-posed formulation (resolved in Sec.~\ref{03}), and `$\triangle$' indicates that the issue remains open.
    }
    \begin{tabular}{c || c | c | c | c}
        Theory Name  
        & Formulation 
        & Configuration variable 
        & Gauge choice
        & Well-posed formulation 
        \\ \hline\hline
        GR 
        & Spacetime 
        & $g_{\mu\nu}$ 
        &
        None (not fixed)
        & $\checkmark$ 
        \\ \hline
        Covariant TEGR 
        &
        Internal-space 
        & 
        $e^{I}{}_{\mu}$, $\omega^{I}{}_{J\mu}$ 
        & 
        None (not fixed)
        & 
        $\checkmark$ 
        \\ \hline
        TEGR
        & Internal-space 
        & $e^{I}{}_{\mu}$ 
        & 
        \begin{tabular}{c}
        Weitzenb\"{o}ck gauge
        \\
        ($\omega^{I}{}_{J\mu}=0$)
        \end{tabular}
        & $\checkmark$ 
        \\ \hline
        Covariant STEGR 
        & 
        Spacetime 
        & 
        $g_{\mu\nu}$, $\Gamma^{\rho}_{\mu\nu}$ 
        & None (not fixed)
        & 
        $\checkmark$ 
        \\ \hline
        CGR 
        & Spacetime 
        & $g_{\mu\nu}$ 
        & \begin{tabular}{c}
        Coincident gauge
        \\
        ($\Gamma^{\rho}_{\mu\nu}=0$)
        \end{tabular}
        & $\times$ (\textit{resolved in Sec.~\ref{03}}) 
        \\ \hline
        Internal STEGR 
        & Internal-space 
        & $g_{\mu\nu}$, $g_{IJ}$, $\xi^{I}$ 
        & 
        None (not fixed)
        & $\triangle$ 
        \\ \hline
    \end{tabular}
    \label{table: Theories and status in first group issue}
\end{table}

\section{\label{02}Gravity in Flat and Torsion-Free Spacetime}

\subsection{\label{02:01}Formulations of Symmetric Teleparallel Equivalent to GR}

In MAG, GR has two equivalent classes of descriptions employing action integrals~\cite{BeltranJimenez:2019esp}. 
One class is TEGR~\cite{Einstein1928}, where the torsion describes gravity,
while another is STEGR~\cite{Nester:1998mp}, where the non-metricity describes gravity. 
STEGR plays the leading role in the current work, and we introduce STEGR in this section (for TEGR, see Ref.~\cite{Bahamonde:2021gfp}).
To restrict generic MAG to STEGR, we impose two conditions: the teleparallel condition (or teleparallelism for short) to make the spacetime flat
\begin{align}
    R^{\rho}{}_{\lambda\mu\nu} 
    := 
    0
\label{teleparallelism}
\end{align}
and the torsion-free condition, 
\begin{align}
    T^{\rho}{}_{\mu\nu} 
    := 
    0
    \,,
\label{torsion-free}
\end{align}
And, there are two methods to implement these conditions: 
\begin{enumerate}
    \item Palatini approach to gravity (Spacetime formulation): \\
    This approach describes gravity in terms of the metric and the affine connection independently. 
    We briefly review the STEGR in the Palatini approach in this section, and the general properties and formulation are discussed in Ref.~\cite{Olmo:2011uz}. 
    \item Gauge approach to gravity (Internal-space formulation): \\
    This approach describes gravity in terms of the internal metric, the connection 1-form, and the frame field, independently. 
    We analyze the STEGR in the Gauge approach in the next sections.  
\end{enumerate}
In the Dirac-Bergmann (DB) analysis~\cite{Dirac:1950pj, Dirac:1958sc, Anderson:1951ta, Bergmann:1949zz, Bergmann1950, BergmannBrunings1949, Sugano:1982bm, Sugano:1989rq, Sugano:1991ir}, which guarantees the consistency of the theory from the point of view of symmetries, the Palatini and gauge approaches are realized by imposing a set of appropriate condition as \textit{outer} primary constraint and \textit{inner} primary constraint, respectively. 
In this work, we do not step into the details of these methods (see Refs.~\cite{Obukhov:2002tm, Ong:2017xwo} and Refs.~\cite{Hehl:1994ue, JimenezCano:2021rlu} for further information on the treatment of outer constraint and inner constraint into the theory).
However, we emphasize that the teleparallelism and torsion-free conditions can be consistently incorporated into the DB analysis.

In this work, we use only the following facts.
The configuration space of the Palatini approach is spanned by the spacetime metric $g_{\mu\nu}$ and affine connection $\Gamma^{\rho}_{\mu\nu}$ as in $Q_{\rm Palatini\,approach} = \left< g_{\mu\nu}\,,\Gamma^{\rho}_{\mu\nu} \right>$.
In contrast, the configuration space of the gauge approach formulation is spanned by the internal metric $g_{I J}$, connection 1-form (components) $\omega^{I}{}_{J\mu}$, and co-frame field (components) $e^{I}{}_{\mu}\,$ as in
$Q_{\rm Gauge\,approach} = \left<g_{I J}
\,,\omega^{I}{}_{J\mu}\,,e^{I}{}_{\mu} \right>$.
We note that the internal metric $g_{I J}$ behaves as a dynamical field rather than a fixed flat internal metric.
One can also find a similar usage of the internal metric in Ref.~\cite{Capozziello:2022zzh}.

\subsection{Symmetric Teleparallel Equivalent to GR in Palatini approach}
\label{02:02}

First, we provide a brief review of the first method for introducing STEGR.
In the configuration space $Q_{\rm Palatini\,approach} = \left< g_{\mu\nu}\,,\Gamma^{\rho}_{\mu\nu}\,;\,\tau_{\rho}{}^{\lambda\mu\nu}\,,\tau_{\rho}{}^{\mu\nu} \right>$, the action functional is given by
\begin{align}
\begin{split}
    &I_{\rm STEGR}[g^{\mu\nu}\,,\Gamma^{\rho}_{\mu\nu}\,,\tau_{\rho}{}^{\lambda\mu\nu}\,,\tau_{\rho}{}^{\mu\nu}] 
    \\
    & \quad
    = 
    \int_{\mathcal{M}}\,
        \left[\,\mathscr{L}_{\rm STEGR}(g_{\mu\nu}\,,Q_{\rho}{}^{\mu\nu}) 
        + \tau_{\rho}{}^{\lambda\mu\nu}\,R^{\rho}{}_{\lambda\mu\nu} 
        + \tau_{\rho}{}^{\mu\nu}\,T^{\rho}{}_{\mu\nu}\,
    \right]
    \,\sqrt{-g}\,dx^{n+1}
    \,,
\end{split}
\label{AI of STEGR}
\end{align}
where $\tau_{\rho}{}^{\lambda\mu\nu}$ and $\tau_{\rho}{}^{\mu\nu}$ are Lagrange multipliers with respect to the generic curvature and the torsion, respectively.
Applying the variation principle with respect to $\tau_{\rho}{}^{\lambda\mu\nu}$ and $\tau_{\rho}{}^{\mu\nu}$, the teleparallelism and torsion-free condition are satisfied, and MAG is now restricted to STEGR theories. 
The field equations are derived as follows:
\begin{align}
\begin{split}
    &\nabla_{\lambda}\left(\sqrt{-g}q^{\lambda}{}_{\mu\nu}\right) 
    + \frac{\delta\left(\sqrt{-g}\mathscr{L}_{\rm STEGR}\right)}{\delta g^{\mu\nu}} 
    = 0
    \,,\\
    &\nabla_{\lambda}\left(\sqrt{-g}\,\tau_{\alpha}{}^{\gamma\beta\lambda}\right) 
    + \left(\sqrt{-g}\,\tau_{\alpha}{}^{\beta\gamma}\right) 
    - \left(\sqrt{-g}\,q^{\beta\gamma}{}_{\alpha}\right) 
    = 0 
    \,,
\end{split}
\label{EoMs of STEGR wrt g and Gamma}
\end{align}
where $q^{\rho}{}_{\mu\nu} := \delta(\mathscr{L}_{\rm STEGR})/\delta Q_{\rho}{}^{\mu\nu}\,$. 
The field equation with respect to the affine connection turns into one with respect to the multipliers. 
We start with the \textit{mimetic} Einstein-Hilbert Lagrangian, $\mathscr{L}_{\rm mimetic\ EH}(g_{\mu\nu}\,,Q_{\rho}{}^{\mu\nu}) = \overset{\circ}{R}$, for the choice of $\mathscr{L}_\mathrm{STEGR}$.
Note that the spacetime manifold $\mathcal{M}$ is non-Riemannian, while GR is formulated in the pseudo-Riemannian manifold. 
The spacetime manifold in STEGR is different from that of GR, and the word `mimetic' emphasizes such a difference.

Applying Eq.~\eqref{generic scalar curvature} to the mimetic Einstein-Hilbert Lagrangian, we obtain
\begin{align}
\begin{split}
    &\mathscr{L}_{\rm STEGR}(g_{\mu\nu}\,,Q_{\rho}{}^{\mu\nu}) 
    \\
    & \quad
    = 
    - \frac{1}{4}Q_{\rho\mu\nu}Q^{\rho\mu\nu} 
    + \frac{1}{2}Q_{\rho\mu\nu}Q^{\mu\nu\rho} 
    - \frac{1}{2}Q_{\mu}\bar{Q}^{\mu} 
    + \frac{1}{4}Q_{\mu}Q^{\mu} 
    + \overset{\circ}{\nabla}{}_{\mu}\left(\bar{Q}^{\mu} - Q^{\mu}\right)
    \,,
\end{split}
\label{mimetic EH Lagrangian}
\end{align}
where $Q_{\mu} = Q_{\mu\nu}{}^{\nu}$ and $\bar{Q}_{\mu} = Q_{\nu\mu}{}^{\nu}$. 
The last term on the right-hand side above is a boundary term and does not affect the field equations. 
Thus, we remove the boundary term in many cases.
We remark that in the above formulation, the condition of vanishing affine connection (which is \textit{not} the coincident gauge) does not generally make the generic curvature vanish in the entire spacetime region.
For a coordinate transformation, $x^{\mu} \rightarrow x'^{\mu} = x'^{\mu}(x^{\mu})\,$, the affine connection is transformed by Eq.~\eqref{Law of gauge transformation of affine connection}.

Taking a specific coordinate system, 
$x^{\rho} \rightarrow x'^{\rho} = x^{\rho} + A^{\rho}_{\mu\nu}\,x^{\mu}\,x^{\nu}/2$  
where $A^{\rho}_{\mu\nu}$ is an arbitrary coefficient being symmetric with respect to the two lower indices, we can prove the proposition that \textit{the affine connection vanishes if and only if the two lower indices are symmetric}. 
Thus, in a local region where the coordinate system $x'$ exists, the torsion-free property leads to the condition of vanishing affine connection,
\begin{align}
    \Gamma^{\rho}_{\mu\nu} = 0
    \,.
\label{mimetic CG in Palatini approach}
\end{align}
We can always choose the affine connection to vanish locally in torsion-free spacetime. 
As a result, the generic curvature also vanishes in the local region where the affine connection vanishes:
\begin{align}
    \sqrt{-g}
    \,R^{\alpha}{}_{\beta\mu\nu} = 
    \sqrt{-g}
    \,\left[2\partial_{[\mu}\Gamma^{\alpha}_{\nu]\beta} + 2\Gamma^{\alpha}_{[\mu|\lambda|}\Gamma^{\lambda}_{\nu]\beta}\right] \underset{\rm x'-coordinate\,system}{=} 
    0 \,.
\label{curvature in component form with mimetic C-gauge}
\end{align}
However, this property does not generally hold in another region where the $x'$-coordinate system cannot cover.\footnote{
For a given manifold $\mathcal{M}$, we can take a chart (coordinate system $x'$ in the manuscript) only in a local open region $U' \subset \mathcal{M}$. 
In such a region $U'$, Eq.~\eqref{mimetic CG in Palatini approach} holds and leads to Eq.~\eqref{curvature in component form with mimetic C-gauge}. 
However, since $\Gamma^{\rho}_{\mu\nu}$ is not a tensor field, this statement does not hold for another chart (say, coordinate system $x''$) covering another local open region $U'' \subset \mathcal{M}$ such that $U' \cap U''$ is not empty. 
Therefore, Eq.~\eqref{curvature in component form with mimetic C-gauge} does not hold in the local closed region $U'' \cap ^\lnot(U' \cap U'') \subset \mathcal{M}\,$ or $U' \cap ^\lnot(U' \cap U'') \subset \mathcal{M}\,$; that is, Eq.~\eqref{curvature in component form with mimetic C-gauge} does not hold globally in $\mathcal{M}$. 
}
To realize teleparallelism, we must impose the property of vanishing generic curvature in addition to that of torsion.\footnote{
Of course, once we obtain a flat and torsion-free spacetime, we can take Eq.~\eqref{mimetic CG in Palatini approach} to hold on the entire spacetime region, including boundaries. 
The crucial point here is that the converse is not true.}

For instance, let us consider formulating a gravity theory on a two-dimensional topological space of M\"{o}bius strip. 
This geometry can be constructed using a vector bundle $(E,S^{1},\pi)$ with a structure group $\mathbb{Z}_{2} = \{-1,1\}$, where $\pi$ is the projection map from $E$ to $S^{1}$.
We cover $S^{1}$ by three open sets: $S^{1} \subset U_{1}\cup U_{2}\cup U_{3}$. 
For each open set, we introduce $\phi_{i}:\pi^{-1}(U_{i})\rightarrow U_{i}\times\mathbb{R}^{2}$, 
which serves as a \textit{local} coordinate function on the total space $E$. 
Then, we define the transition functions by
$\phi_{1}\circ\phi^{-1}_{2} = 1$, $\phi_{1}\circ\phi^{-1}_{3} = 1$, and $\phi_{2}\circ\phi^{-1}_{3} = -1$.
With these transition functions, the total space $E=\sqcup_{i=1}^{3}\pi^{-1}(U_{i})$ becomes the M\"{o}bius strip. 
If we consider an action integral on $\pi^{-1}(U_{1})$ with a constraint to vanish the curvature by using a Lagrange multiplier, $\pi^{-1}(U_{1})$ becomes a flat 2-dimensional rectangle.
However, on $\pi^{-1}(U_{2})$ and $\pi^{-1}(U_{3})$, this optimization problem admits no solution because of the twist inherent in the M\"{o}bius strip.
Thus, imposing \textit{a priori} that the spacetime ($S^{1}$ here) is globally flat is inconsistent.

The above example illustrates a pathology that always appears in the Palatini approach to teleparallel theories of gravity: 
global flatness cannot be consistently imposed through Lagrange multipliers on a non-trivial bundle.

\section{Symmetric Teleparallel Equivalent to GR in Internal-Space Formulation}
\label{03}

\subsection{Internal STEGR in Formalism 3 and its field equations}
\label{03:01}

In STEGR, the action integral is formulated using the Lagrange multiplier method, as first attempted in Sec.~\ref{02:01}. 
On the other hand, the second method explained in Sec.~\ref{02:01} can be applied with the so-called Weitzenb\"{o}ck gauge to TEGR only since in this gauge both the generic curvature and the non-metricity automatically vanish, and the torsion survives only~\cite{Ferraro:2016wht, Tomonari:2024vij}. 
Such a structure arises because the local Lorentz symmetry restricts the possible geometric quantities that exist in the geometry.
In the Weitzenb\"{o}ck gauge, it is impossible to formulate STEGR since the non-metricity is a tensor field in this theory and, thus, vanishing it in a certain coordinate system results in it vanishing in any coordinate system.

Ref.~\cite{Tomonari:2024vij} reformulated the theories of STEGR by generalizing the internal-space bundle so that the non-metricity arises from the internal-space structure by extending/reducing the local Lorentz symmetry to a more generic one. 
Consequently, the internal-space metric alters the Minkowski metric to a generic one. 
Moreover, Ref.~\cite{Tomonari:2024vij} proposed three possible formalisms of STEGR as follows.
Formalism 1 is essentially the same as one introduced in Sec.~\ref{02:02}, while Formalisms 2 and 3 are completely new.
Formalism 3 resolves the inadequate use of the coincident gauge and introduces CGR~\cite{BeltranJimenez:2017tkd} in a well-posed manner.
It is the first attempt to introduce CGR in the Weitzenb\"{o}ck gauge without any issue. 
Note that there are other formalisms; for instance, Ref.~\cite{Hu:2023gui} partially utilizes the internal-space formalism.
Ref.~\cite{Hu:2023gui} also introduces a Lagrange multiplier to reformulate the nonlinear extension of STEGR, covariant $f(Q)$ theory, as the scalar-vector-tensor theory.

In this subsection, we briefly review Formalism 3 in Ref.~\cite{Tomonari:2024vij}.
We start with the affine connection in the Weitzenb\"{o}ck gauge introduced by Eq.\eqref{Weitzenboeck gauge} in Sec.~\ref{addedSec}. 
The result is
\begin{align}
    \overset{\rm w}{\Gamma}{}^{\rho}_{\mu\nu} = e_{I}{}^{\rho}\,\partial_{\mu}\,\theta^{I}{}_{\nu}\,,
\label{Weitzenboch connection}
\end{align}
where $e_{I}{}^{\mu}$ and $\theta^{I}{}_{\mu}$ are the frame field and co-frame field component. 
A straightforward calculation with Eq.~\eqref{Weitzenboch connection} shows \textit{a priori} satisfaction of vanishing generic curvature in action integrals by its definition as long as we assume the Weitzenb\"{o}ck gauge.
That is, we can regard it as $R^{\alpha}{}_{\beta\mu\nu} = 0\,$ in the entire spacetime region, which differs from the case of the Palatini approach.
All algebraic terms are canceled out in the use of $e_{I}{}^{\mu}e^{J}{}_{\mu} = \delta_{I}{}^{J}\,$, $e_{I}{}^{\mu}e^{I}{}_{\nu} = \delta^{\mu}{}_{\nu}$, and the Leibniz rule shows that, in total, the generic curvature vanishes up to boundary terms under the assumption of the Weitzenb\"{o}ck gauge. 

Explicitly, we obtain\footnote{
We can show Eq.~\eqref{special boundary term in STEGR} as follows:
\begin{align}
\begin{split}
    \sqrt{-g}\,R^{\alpha}{}_{\beta\mu\nu} &= \sqrt{-g}\,\left[2e_{I}{}^{\alpha}e^{J}{}_{\beta}\partial_{[\mu|}\omega^{I}{}_{J|\nu]} + 2\partial_{[\mu}\overset{\rm w}{\Gamma}{}^{\alpha}_{\nu]\beta} + 2\overset{\rm w}{\Gamma}{}^{\alpha}_{[\mu|\lambda|}\overset{\rm w}{\Gamma}{}^{\lambda}_{\nu]\beta}\right] \\
    &= \partial_{[\mu|}\left(2\,\sqrt{-g}\,e_{I}{}^{\alpha}e^{J}{}_{\beta}\omega^{I}{}_{J|\nu]}\right) - \partial_{[\mu|}\left(2\,\sqrt{-g}\,e_{I}{}^{\alpha}e^{J}{}_{\beta}\right)\omega^{I}{}_{J|\nu]} + \sqrt{-g}\,\left[2\partial_{[\mu}e_{I}{}^{\alpha}\partial_{\nu]}\theta^{I}{}_{\beta} + 2e_{I}{}^{\alpha}\partial_{[\mu}\theta^{I}{}_{|\lambda|}e_{J}{}^{\lambda}\partial_{\nu]}\theta^{J}{}_{\beta}\right] \\
    &= \partial_{[\mu|}\left(2\,\sqrt{-g}\,e_{I}{}^{\alpha}e^{J}{}_{\beta}\omega^{I}{}_{J|\nu]}\right) + \sqrt{-g}\,\left[2\partial_{[\mu}e_{I}{}^{\alpha}\partial_{\nu]}\theta^{I}{}_{\beta} - 2e_{I}{}^{\alpha}\theta^{I}{}_{\lambda}\partial_{[\mu}e_{J}{}^{\lambda}\partial_{\nu]}\theta^{J}{}_{\beta}\right] \\
    &= \partial_{[\mu|}\left(2\,\sqrt{-g}\,e_{I}{}^{\alpha}e^{J}{}_{\beta}\omega^{I}{}_{J|\nu]}\right)\,.
\end{split}
\end{align}
In the third line, we imposed the Weitzenb\"{o}ck gauge. 
In the third and the fourth lines, we used $e_{I}{}^{\mu}e^{J}{}_{\mu} = \delta_{I}{}^{J}$ and $e_{I}{}^{\mu}e^{I}{}_{\nu} = \delta^{\mu}{}_{\nu}$, respectively. 
}
\begin{align}
    \sqrt{-g}
    \,R^{\alpha}{}_{\beta\mu\nu} = \partial_{[\mu|}\left(2\,
    \sqrt{-g}
    \,e_{I}{}^{\alpha}e^{J}{}_{\beta}\omega^{I}{}_{J|\nu]}\right)
    \,.
\label{special boundary term in STEGR}
\end{align}
Thus, this boundary term vanishes as far as we impose the Weitzenb\"{o}ck gauge. 
We emphasize that the way to vanish the generic curvature given in Eq.~\eqref{decomposition of curvature} differs from that in the Palatini approach.
Eq.~\eqref{curvature in component form with mimetic C-gauge} is valid only in a local region such that the ``$x'$-coordinate system'' exists. 
In a global region including a boundary, such a coordinate system does not generally exist.
Whereas, since the connection 1-form is defined in a coordinate-independent manner, the Weitzenb\"{o}ck gauge holds at every spacetime point.
We note that in the internal space, the generic curvature does not vanish: 
$R = d_{\nabla}\omega = d\omega\,$, or equivalently, $R^{I}{}_{J\mu\nu} = 2\partial_{[\mu|}\omega^{I}{}_{J|\nu]}\,$. 
That is, Eq.~\eqref{special boundary term in STEGR} is nothing but the push forward of the generic curvature in the internal space.

Next, we impose the co-frame field decomposition introduced by Eq.~\eqref{Stueckelberg field} in Sec.~\ref{addedSec}. 
In addition to this decomposition, we impose
\begin{align}
    \xi^{I}\partial_{\mu}g_{IJ} := 0
    \,.
\label{nec and suf cond for e theta = delta}
\end{align}
If the local symmetry is Lorentzian, the above condition is automatically satisfied.
However, it is not the case in our current consideration since the decomposition of $\theta^{I}{}_{\mu} = \partial_{\mu}\xi^{I}$ breaks the local Lorentz symmetry. 
This condition leads to the convenient formula given as follows:
\begin{align}
    \partial_{\mu}\xi_{A} = \eta_{AB}\,\partial_{\mu}\xi^{B}
    \,.
\end{align}
Then, the affine connection and the action integral are provided as follows~\cite{BeltranJimenez:2017tkd, Tomonari:2024vij}:
\begin{align}
    \overset{s}{\Gamma}{}^{\rho}_{\mu\nu} =: \frac{\partial x^{\rho}}{\partial \xi^{I}}\,\partial_{\mu}\,\partial_{\nu}\,\xi^{I}\,
\label{Weitzenboch connection in Stuckelberg decomposition}
\end{align}
and
\begin{align}
\begin{split}
    &I_{\rm\,internal\,STEGR\,in\,Formalism\,3\,}[\xi_{A}{}^{}\,,g^{\mu\nu}\,,g_{AB}] 
    \\
    & \quad
    = 
    \int_{\mathcal{M}}dx^{4}\,\frac{1}{2}\,\xi\,\sqrt{-\mathrm{det}(g_{A B})}\,g^{ABCDEF}\,g^{\mu\alpha}\,g^{\nu\beta}\,\partial_{\alpha}\xi_{A}{}^{}\,\partial_{\beta}\xi_{B}{}^{}\,\partial_{\mu}g_{C[D|}\partial_{\nu}g_{|E]F}
    \,, 
\end{split}
\label{STEGR Lagrangian in terms of xi and eta in Formalism 3}
\end{align}
where
\begin{align}
    \quad g^{ABCDEF} := g^{AB}g^{CD}g^{EF} + 2g^{AF}g^{BD}g^{CE}
    \,.
\end{align}
Here, $\xi$ is the determinant of $\partial^{\mu}\xi_{A} = g^{\mu\nu}\,\partial_{\nu}\xi_{A}\,$, $g^{IJ}$ is the inverse of the internal-space metric tensor (component) $g_{IJ}$. 
Note that the configuration space is now 
$Q_{\rm Gauge\,approach} = \left< g_{\mu\nu}\,;\,g_{IJ}\,,\xi^{I} \right>\,$. 

At first glance, the above bi-metric structure suggests some incompleteness of the theory in the gauge approach. 
However, we will see that this incompleteness does not affect the derivation of CGR. 
From the Lagrangian above, we find that the internal STEGR in Formalism 3 is a diffeomorphism-invariant theory.\footnote{
Under the imposition of the Weitzenb\"{o}ck gauge, we remark that the covariant derivative acting on a quantity only with internal-space indices results in the partial derivative.
}

The decomposition of $\theta^{I}{}_{\mu} = \partial_{\mu}\xi^{I}$ implies the torsion-free condition: $T = de^{-1} = 0\,$, or equivalently, $T^{I}{}_{\mu\nu} = 2\partial_{[\mu}\theta^{I}{}_{\nu]} = 0$, suggesting that the torsion also vanishes in spacetime: $T^{\rho}{}_{\mu\nu} = 0\,$. 
We remark that this result is consistent with that of the Palatini approach, thanks to Eq.~\eqref{Weitzenboch connection in Stuckelberg decomposition}.
The non-metricity, $Q = d\eta$, or equivalently, $Q_{\mu IJ} = \partial_{\mu}g_{IJ}$, becomes
\begin{align}
    \overset{\rm s}{Q}{}_{\rho\mu\nu} =: \partial_{\mu}\xi^{I}\,\partial_{\nu}\xi^{J}\,\partial_{\rho}g_{IJ}
    \,.
\label{non-metricity in internal STEGR in Formalism 3}
\end{align}
In the existing works, the internal metric $g_{IJ}$ is always fixed as the Minkowski one, and the non-metricity and the Lagrangian always vanish whether STEGR is on the spacetime or internal space or not. 
Note that the above bi-metric structure is unique to STEGR and different from the known structure in the bigravity~\cite{Rosen:1940zza, Rosen:1940zz, Hassan:2011zd, Schmidt-May:2015vnx}. 
We also emphasize that the decomposition in Eq.~\eqref{Weitzenboch connection in Stuckelberg decomposition} implies the introduction of a global symmetry, which reduces the apparent DOFs of the theory~\cite{Tomonari:2024vij}.

The field equations of the internal STEGR in Formalism 3 are derived as follows:
\begin{align}
\begin{split}
    &F^{(0)\,A}{}_{B}{}^{|\alpha\beta}\partial_{\alpha}\partial_{\beta}\xi^{B} 
    + F^{(1)}\,g^{\alpha\beta}\,\partial_{\alpha}\partial_{\beta}\xi^{A} 
    + F^{(2)\,A}{}_{BC}{}^{|\mu\nu\rho}\,\partial_{\mu}g^{BC}{}_{\nu\rho} 
    + F^{(3)}{}_{B}{}^{|\mu\nu\rho}\,\partial_{\mu}\tilde{g}
    ^{AB}{}_{\nu\rho} 
    + F^{(4)\,A} 
    = 0 
    \,,\\
    &G^{(0)\,ABCD|\mu\nu}\,\partial_{\mu}\partial_{\nu}g_{CD} 
    + G^{(1)\,ABC|\mu\nu}\,\partial_{\mu}\partial_{\nu}\xi_{C} 
    + G^{(2)\,ABC}\,g^{\alpha\beta}\partial_{\alpha}\partial_{\beta}\xi_{C} 
    + G^{(3)\,AB} 
    = 0 
    \,,\\
    &g^{\alpha\beta}\tilde{g}^{AB}{}_{\mu\alpha}\partial_{\nu}\xi_{A}\partial_{\beta}\xi_{B} 
    = 0
    \,,
\end{split}
\label{Field eqs of internal STEGR in formalism 3}
\end{align}
where $\eta^{AB}{}_{\mu\nu}$ and $\tilde{\eta}^{AB}{}_{\mu\nu}$ are set as 
\begin{align}
\begin{split}
    &g{}^{AB}{}_{\mu\nu} 
    := 
    \sqrt{-\mathrm{det}(g_{AB})}g^{ABCDEF}g_{CDEF|\mu\nu}
    \,,\\
    &
    g_{CDEF|\mu\nu} 
    := 
    \partial_{\mu}g_{C[D|}\partial_{\nu}g_{|E]F}
    \,,\\
    &\tilde{g}^{AB}{}_{\mu\nu} 
    := 
    g{}^{AB}{}_{\mu\nu} + g{}^{BA}{}_{\nu\mu} 
    = 
    \sqrt{-\mathrm{det}{(g_{AB})}}\,g^{ABCDEF}\tilde{g}_{CDEF|\mu\nu} 
    = 
    \tilde{g}^{BA}{}_{\nu\mu}
    \,,\\
    &\tilde{g}_{CDEF|\mu\nu} 
    := 
    \partial_{\mu}g_{CD}\partial_{\nu}g_{EF} - \partial_{(\mu|}g_{CE}\partial_{|\nu)}g_{DF} 
    = 
    \tilde{g}_{DCFE|\mu\nu} 
    = 
    \tilde{g}_{EFCD|\nu\mu}
    \,.
\end{split}
\label{eta and tidle_eta}
\end{align}
Explicit expressions of $F^{(a)}$ $(a = 0\,,1\,,2\,,3\,,4)$ and $G^{(i)}$ $(i = 0\,,1\,,2\,,3)$ (abbreviated indices here) are given in Appendix~\ref{app:01}. 
Notably, these entities are composed only of up to first-order partial derivatives of $\xi_{I}$, $g_{\mu\nu}$, and $g_{IJ}$.
In applying the variational principle, we assume the Dirichlet boundary conditions, $\delta\xi_{A}|_{\partial\mathcal{M}} \underset{\rm set}{:=} 0$ and $\delta g^{AB}|_{\partial\mathcal{M}} := 0$.
Note that we do not need to impose any boundary condition on the spacetime metric such as $\delta g^{\mu\nu}|_{\partial\mathcal{M}} \underset{\rm set}{:=} 0\,$ to derive the third field equations. 
It implies that in this theory, the spacetime metric 
has the same property of Lagrange multiplier.

Due to the analytical complexity of the derived field equations~\eqref{Field eqs of internal STEGR in formalism 3}, the present work focuses on examining their general structure in relation to ghost DOFs.
Since the first equation in Eq.~\eqref{Field eqs of internal STEGR in formalism 3} contains the second-order terms of both $\xi_{A}$ (or equivalently, $\xi^{A}\,$,) and 
$g_{IJ}$, these variables should be dynamical.
The second equation in Eq.~\eqref{Field eqs of internal STEGR in formalism 3} has the same structure as the first equation, which may raise the issue of the existence of ghost DOFs.
That is, $\xi_A$ and $g_{IJ}$ can be expressed in equations involving derivatives of up to fourth order.
To clarify this point, let us consider the following conditions:
\begin{align}
\begin{split}
    &F^{(0)\,A}{}_{B}{}^{|\alpha\beta}\,\tilde{F}^{(0)}{}_{A} 
    \underset{\rm if\,\it\tilde{F}^{(0)}{}_{A}\rm\,exist}{=} 
    - F^{(1)}\,g^{\alpha\beta}\,\tilde{F}^{(0)}{}_{B} 
    = 
    - F^{(1)}\,g^{\alpha\beta}\,\delta^{A}{}_{B}\,\tilde{F}^{(0)}{}_{A}
    \,,\\
    &G^{(1)\,ABC|\mu\nu}\,\tilde{G}^{(1)}{}_{AB} 
    \underset{\rm if\,\it\tilde{G}^{(1)}{}_{AB}\rm\,exist}{=} 
    -G^{(2)\,ABC}\,g^{\mu\nu}\,\tilde{G}^{(1)}{}_{AB}
    \,,
\end{split}
\label{conditions to make field equs of internal STEGR in formalism 3 simplier}
\end{align}
for the assumptions of existing non-vanishing $\tilde{F}^{(0)}{}_{A}$ and $\tilde{G}^{(1)}{}_{AB}$. 
The first and second equations in Eq.~\eqref{Field eqs of internal STEGR in formalism 3} turn out to be
\begin{align}
\begin{split}
    &\tilde{F}^{(0)}{}_{A}
    \left(
        F^{(2)\,A}{}_{BC}{}^{|\mu\nu\rho}\,\partial_{\mu}g^{BC}{}_{\nu\rho} 
        + F^{(3)}{}_{B}{}^{\mu\nu\rho}\,\partial_{\mu}\tilde{g}^{AB}{}_{\nu\rho} 
        + F^{(4)\,A}
    \right) 
    = 0
    \,,\\
    &\tilde{G}^{(1)}{}_{AB}
    \left(
        G^{(0)\,ABCD|\mu\nu}\,\partial_{\mu}\partial_{\nu}g_{CD} + G^{(3)\,AB}
    \right) 
    = 0
    \, .
\end{split}
\label{Simpler field eqs of internal STEGR in formalism 3}
\end{align}
Although the above equations seem to give rise to the first-order derivative of variables $\xi_{A}$, one would verify that they include at most third-order derivatives of $\xi_{A}$.
Let us assume that a domain exists for $g_{\mu\nu}$ and $\xi_{A}$ such that the second equation in Eq.~\eqref{Simpler field eqs of internal STEGR in formalism 3} has a solution. 
We denote the solution by 
$g_{IJ} = g_{IJ}(g_{\mu\nu}\,,\partial_{\rho}g_{\mu\nu}\,;\,\xi_{A}\,,\partial_{\mu}\xi_{A})\,$. 
Substituting this solution into the first equation in Eq.~\eqref{Simpler field eqs of internal STEGR in formalism 3}, we can verify that the equations have the derivative terms of $\xi_{A}$ up to third-order. 
Thus, there would exist ghost DOFs in STEGR. 
However, in terms of field equations, we cannot proceed further to unveil the existence of ghost DOFs.

This situation suggests the necessity of performing the DB analysis and imposing appropriate degenerate conditions~\cite{Motohashi:2016ftl}. 
In this theory, the DB analysis can avoid the emergence of unsolvable partial differential equations~\cite{DAmbrosio:2023asf} since the theory possesses diffeomorphism invariance.
That is, taking ADM-foliation properly, we can avoid the problematic spatial boundary terms. 
One can use this prescription without loss of generality thanks to the diffeomorphism invariance. 
In detail, see Ref.~\cite{Tomonari:2023wcs}.
We note that a bifurcation would occur due to the violation of the local Lorentz symmetry in this analysis~\cite{Tomonari:2023wcs}. 
Another possibility for remedying the pathology of Ostrogradski's ghost is to adjust the fundamental variables spanning the configuration space. 
A possible choice of the fundamental variables is proposed in Ref.~\cite{Hu:2023gui},
and the same variable choice is also discussed within the internal STEGR based on  Formalism 2~\cite{Tomonari:2024vij}. 
We will consider this issue together with revealing the true nature of the bi-metric structure.

\subsection{Revisiting Coincident GR}
\label{03:02}

Motivated by the considerations in the previous subsection, we now formulate CGR based on the internal STEGR in Formalism 3.
First, the local structure $\left.\mathcal{M}\times\mathbb{R}^{n+1}\right|_{U}\simeq \left.T\mathcal{M}\right|_{U}$ allows us to identify the internal-space indices with the spacetime indices at least on a local region $U$: $\mu\,,\nu\,,\rho\,,\cdots = I\,,J\,,K\,,\cdots\,$. 
This local geometric property allows us to apply the coincident gauge, which is introduced by Eq.~\eqref{coincident gauge in generic form} in Sec.~\ref{addedSec}. That is, 
\begin{align}
    \xi^{I} := \delta^{I}{}_{\mu}x^{\mu}\,
\label{coincident gauge}
\end{align}
for all charts in an atlas, where we set $a^{\mu} = 0$ since this term does not affect the result. 
Here, the generic Kronecker's delta $\delta^{I}{}_{\mu}$ takes its value $1$ for $I = \mu$ as an integer running from $0$ to the dimension of space and $0$ for other combinations. 
$\delta_{I}{}^{\mu}$ is defined in the same way. 
This means $U := \mathcal{M}$, which is the globalization of $\left.\mathcal{M}\times\mathbb{R}^{n+1}\right|_{U}\simeq \left.T\mathcal{M}\right|_{U}$. 
Consequently, we obtain a set of fundamental formulae:
\begin{align}
\begin{split}
    &\eta_{IJ} 
    \underset{\rm C-gauge}{=} 
    \delta^{\mu}{}_{I}\,\delta^{\nu}{}_{J}\,g_{\mu\nu}
    \,,\\
    &\partial_{\alpha}\xi_{A} 
    \underset{\rm C-gauge}{=} 
    \delta_{A}{}^{\beta}g_{\alpha\beta}
    \,,\\
    &\partial_{\alpha}\xi^{A} 
    \underset{\rm C-gauge}{=} 
    \delta^{A}{}_{\alpha}
    \,,
\end{split}
\label{fundamental algebra from internal STEGR in formalism 3 to CGR}
\end{align}
where `C-gauge' is the abbreviation of `Coincident gauge'. 
Eqs.~\eqref{Weitzenboch connection in Stuckelberg decomposition} and \eqref{non-metricity in internal STEGR in Formalism 3} become
\begin{align}
    \overset{\rm c}{\Gamma}{}^{\rho}_{\mu\nu} = 0\,
\label{Affine connection in coincident gauge}
\end{align}
and
\begin{align}
    \overset{\rm c}{Q}{}_{\rho\mu\nu} = \partial_{\rho}g_{\mu\nu}
    \, .
\label{non-metricity in coincident gauge}
\end{align}
We remark that Eq.~\eqref{non-metricity in coincident gauge} is no longer a tensorial quantity. Therefore, a theory composed of this quantity loses the diffeomorphism invariance.

Based on Eq.~\eqref{STEGR Lagrangian in terms of xi and eta in Formalism 3}, the configuration space turns into $Q_{\rm Gauge\,approach} = \left< g_{\mu\nu} \right>\,$, and the action integral becomes as follows:
\begin{align}
\begin{split}
    &I_{\rm CGR}[g^{\mu\nu}] 
    \\
    & \quad= 
    \int_{\mathcal{M}}\,dx^{4}\,\frac{1}{4}\,\sqrt{-g}\,\left(g^{\mu\nu}g^{\rho\sigma}g^{\lambda\kappa} - g^{\mu\nu}g^{\rho\lambda}g^{\sigma\kappa} + 2g^{\mu\kappa}g^{\nu\sigma}g^{\rho\lambda} -2g^{\mu\kappa}g^{\nu\lambda}g^{\rho\sigma}\right)\overset{\rm c}{Q}_{\mu\rho\sigma}\overset{\rm c}{Q}_{\nu\lambda\kappa}
    \,.
\end{split}
\label{AI of CGR}
\end{align}
Eq.~\eqref{AI of CGR} is nothing but the action integral of CGR derived first in Ref.~\cite{Einstein1916} and rediscovered in Ref.~\cite{Runkla:2018xrv} in the context of scalar non-metricity theories. 
We remark that this action is generally defined in a global region $U$,\footnote{
Here, `global' means that there exists an open set $U$ of $\mathcal{M}$ such that any open set $U'$ of $\mathcal{M}$ is contained in $U\,$.
}
which does not always coincide with the spacetime manifold $\mathcal{M}$, such that a regular coordinate system that covers $U$ exists. 
Thus, in general, the action is valid only on $U$. 
Taking into account this precise understanding, let us define a new jargon \textit{C-gauged internal space} by meaning an internal space equipped with a coincident gauge, Eq.~\eqref{coincident gauge}, for a global region $U$. 
If we can take a coordinate system such that no coordinate singularity exists, $U$ coincides with $\mathcal{M}\,$. 
In the following analysis, we assume such a simple case, where coordinate functions can always cover the entire spacetime manifold.

Ref.~\cite{Adak:2008gd} pioneered the above consideration.
In their work, the authors state that the vanishing connection 1-form under ``$e^{\alpha} = dx^{\alpha}$'' in their notation leads to the curvature and the torsion automatically vanishing, and the non-metricity surviving only. 
It seems that both Palatini and gauge approaches are compatible. 
However, there is a trick at this point. 
Ref.~\cite{Adak:2008gd} imposes the condition ``$e^{\alpha} = dx^{\alpha}$'' and the Weitzenb\"{o}ck gauge, but the former condition is nothing more than the coincident gauge. 
Thus, their result is consistent with our work. 
Strictly speaking, the approach in Ref.~\cite{Adak:2008gd} confused the affine connection on a spacetime manifold with the connection 1-form on an internal space. 
If one gets rid of the condition ``$e^{\alpha} = dx^{\alpha}$'', the formulation would encounter an issue. 
We cannot extend the first-order formalism of GR to MAG theories without modifying it appropriately in terms of gauge theories of gravity.
In such a generic case, we should strictly separate and treat each connection~\cite{Tomonari:2023ars}.

Let us consider the consistency of CGR with the internal STEGR in Formalism 3 in terms of the field equation. 
Varying Eq.~\eqref{AI of CGR} with respect to $g^{\mu\nu}$, we obtain the field equations:
\begin{align}
    &h^{\rho\sigma\mu\nu\lambda\kappa}\,\partial_{\rho}\partial_{\sigma}g_{\lambda\kappa} 
    + H^{\mu\nu}(g_{\alpha\beta}\,,\partial_{\gamma}g_{\alpha\beta}) 
    = 0
    \,,
\label{Field eqs of CGR without BTs}
\end{align}
where
\begin{align}
\begin{split}
    &\quad h^{\rho\sigma\mu\nu\lambda\kappa} 
    := 
    -\frac{1}{2}\sqrt{-g}\left\{-g^{\rho\sigma\mu\nu\lambda\kappa} 
    + \frac{1}{2}\left(g^{\rho\sigma\mu\lambda\nu\kappa} 
    + g^{\rho\sigma\nu\lambda\kappa\mu}\right)\right\}
    \,,\\
    &\quad g^{\rho\sigma\mu\nu\lambda\kappa} 
    := 
    g^{\rho\sigma}g^{\mu\nu}g^{\lambda\kappa} 
    + 2g^{\rho\kappa}g^{\sigma\nu}g^{\mu\lambda}
    \, .
\end{split}
\end{align}
The explicit formula of $H^{\alpha\beta} := H^{\alpha\beta}(g_{\mu\nu}\,,\partial_{\gamma}g_{\mu\nu})$ is given in Appendix~\ref{app:02}. 
The important property of $H^{\mu\nu}$ here is that it consists of derivative terms up to first order only. 
In the coincident gauge, performing a tedious but straightforward calculation, we can show the following properties:
\begin{align}
\begin{split}
    &\delta_{A}{}^{\alpha}\,\delta_{B}{}^{\beta}\,G^{(0)\,ABCD|\mu\nu} \underset{\rm C-gauge}{=} h^{\mu\nu\alpha\beta\rho\lambda}\,\delta^{C}{}_{\rho}\,\delta^{D}{}_{\lambda}\,,\\
    &\delta_{A}{}^{\alpha}\,\delta_{B}{}^{\beta}\,\left(G^{(1)\,ABC|\mu\nu}\,\partial_{\mu}\partial_{\nu}\xi_{C} + G^{(2)\,ABC}\,g^{\alpha\beta}\partial_{\alpha}\partial_{\beta}\xi_{C} + G^{(3)\,AB}\right) \underset{\rm C-gauge}{=} H^{\alpha\beta}\,,
\end{split}
\label{internal STEGR in Formalism 3 in C-gauge}
\end{align}
where we used the formulae given in Appendix~\ref{app:03}. 
Using these properties, we can easily derive Eq.~\eqref{Field eqs of CGR without BTs} from the second equation in Eq.~\eqref{Field eqs of internal STEGR in formalism 3}, where we used the formulae given in Eq.~\eqref{fundamental algebra from internal STEGR in formalism 3 to CGR}.

The first equation in Eq.~\eqref{Field eqs of internal STEGR in formalism 3} gives rise to constraints $\phi^{A} \underset{\rm set}{:=}F^{(4)\,A}=0\,$; the first and second terms vanish in the coincident gauge, and the third and fourth terms cancel each other.
The third equation in Eq.~\eqref{Field eqs of internal STEGR in formalism 3} composes a new constraint by 
$\phi_{IJ} := 2\delta_{(I}{}^{\mu}\delta_{J)}{}^{\nu}g^{\alpha\beta}\tilde{g}^{AB}{}_{\mu\alpha}\partial_{\nu}\xi_{A}\partial_{\beta}\xi_{B} = 2g_{A(I}\tilde{g}^{AB}{}_{J)B} = 0\,$
in the internal space. 
These constraints would restrict the configuration space 
$Q_{\rm Gauge\,approach} = \left< g_{\mu\nu}\,;\,g_{IJ}\,,\xi^{I} \right>$ 
to $Q_{\rm Gauge\,approach} = \left< g_{\mu\nu} \right>\,$. 
To verify this, we should perform the DB analysis; however, in the coincident gauge, the theory lacks diffeomorphism invariance. 
Thus, the analysis would suffer from the pathology indicated in Ref.~\cite{DAmbrosio:2023asf}.

Finally, we claim the following proposition: \textit{The field equations of CGR coincide with those of GR up to an overall sign, taking into account the freedom in choosing boundary terms in the variational principle.}
Particularly, the field equation of CGR is entirely equivalent to that of GR in the vacuum case.
Eq.~\eqref{generic scalar curvature} and Eq.~\eqref{decomposition of affine connection} lead to the Lagrangian of CGR:
\begin{align}
    L_{\rm E} = \sqrt{-g}\,\mathscr{L}_{\rm E} := 2\,\sqrt{-g}\,g^{\mu\nu}\,\overset{\circ}{\Gamma}{}^{\rho}_{\lambda[\mu}\overset{\circ}{\Gamma}{}^{\lambda}_{\rho]\nu} = -\,2\,\sqrt{-g}\,g^{\mu\nu}\,\overset{\circ}{\Gamma}{}^{\rho}_{\lambda[\rho}\overset{\circ}{\Gamma}{}^{\lambda}_{\mu]\nu}\,,
\label{Einstein Lagrangian with opposite sign}
\end{align} 
which was first introduced by A. Einstein in 1916~\cite{Einstein1916} (with the opposite sign). 
A straightforward calculation confirms that Eq.~\eqref{Einstein Lagrangian with opposite sign} coincides with the Lagrangian of CGR given in Eq.~\eqref{AI of CGR}. 
In the variational principle, a theory is equivalent to others up to boundary terms. 
Thus, we can freely add a boundary term $\partial_{\rho}(\sqrt{-g}\,w^{\rho})$, where $w^{\rho} \underset{\rm set}{:=} -\,2\,\overset{\circ}{\Gamma}{}^{[\rho}_{\mu\nu}\,g^{\mu]\nu}\,$. 
We obtain
\begin{align}
    L_{\rm opposite\,sign\,mimetic\,EH} := -\,\sqrt{-g}\,\overset{\circ}{R}\,.
\label{E-H Lagranaian with oppisite sign}
\end{align}
Taking into account the Gibbons-York-Hawking term~\cite{Gibbons:1976ue, York:1972sj, Erdmenger:2023hne} with \textit{opposite sign} and varying with respect to the metric tensor under the imposition of the Dirichlet boundary condition, we get
\begin{align}
    -\overset{\circ}{G}_{\mu\nu} = -\,\left(\overset{\circ}{R}{}_{\mu\nu} - \frac{1}{2}\,g_{\mu\nu}\,\overset{\circ}{R}\right) = 0\,.
\label{opposite sign Einstein equation}
\end{align}
In the vacuum case, CGR is completely equivalent to GR. 
Here, taking into account the boundary term in the constitution of the original Lagrangian of CGR given in Eq.~\eqref{Einstein Lagrangian with opposite sign}, we must add the boundary term multiplied by a factor of two. 
We note that, if matter fields are present, the overall sign can be consistently absorbed by multiplying both sides of the field equations by $-1$, so that the physical equations coincide with GR. 
Thus, the opposite sign is merely a convention and does not imply antigravity. 

We remark on our findings.
First of all, we have unveiled that CGR is not a gauge-fixed class of \textit{ordinary} STEGR but that of \textit{internal} STEGR.
Based on the gauge approach to gravity and internal-space formulation, we have clarified the gauge structure associated with the internal space in CGR, although the conventional Palatini approach to STEGR does not manifestly provide it.
By connecting our new perspectives with the results of pioneering works, our novel formulation also redefines the meaning of the `coincident gauge' as the actual gauge fixing in terms of the internal space.
Moreover, regarding the phrase `equivalent to GR', we find that the equivalence is just \textit{formality}. 
For instance, CGR concludes Einstein's equation in the vacuum case as in Eq.~\eqref{opposite sign Einstein equation}, but \textit{the spacetime manifold itself is different from that of GR}. 
That is, the spacetime is a pseudo-Riemannian differential manifold in GR, but it is a flat and torsion-free differential manifold in CGR. 
As a notable point, the Lagrangian in Eq.~\eqref{AI of CGR} is \textit{not} a scalar field.
We find that CGR is not a diffeomorphism-invariant theory without the inclusion of appropriate boundary terms.

We note that ordinary STEGR is formulated on a spacetime manifold and its tangent bundle only, not on a generalized internal space. 
In internal STEGR, the internal space is not restricted to the tangent bundle of the spacetime manifold; the internal bundle is generalized into a more generic one by a bundle morphism in each local region of spacetime. 
This generalization allows us to replace the internal metric with a generic one such that the non-metricity does not vanish even in Weitzenb\"{o}ck gauge on the spacetime manifold~\cite{Tomonari:2024vij}.

\section{Kinematics in Flat and Torsion-Free Spacetime}
\label{04}

Although we have focused on the theoretical aspects of CGR in the previous section, a debatable point remains regarding its physical interpretation. 
The field equation of CGR is completely the same as that of GR in the vacuum case.
However, the description of gravity is entirely different. 
The spacetime of STEGR is a \textit{flat} and \textit{torsion-free} metric-affine differential manifold independent of whether the formulation is ordinary or not.
CGR is nothing more than a gauge-fixed class of \textit{internal} STEGR.  
Thus, even if the field equation coincides with that of GR, the kinematical description of a test particle may differ when the particle action is generalized so that it remains well-defined in the presence of non-metricity.

To study the kinematics in STEGR theories, we first note a potential difficulty concerning the definition of particle mass in the presence of non-metricity~\cite{Wada2023, Wada:2025szz}.
In relativistic theory, the mass is defined as the norm of the four-momentum vector, $m^2 = - p^\mu p_\mu$.
However, in the presence of non-metricity, the norm of a vector changes under parallel transport, leading to a variation of $p^{\mu}$ proportional to the non-metricity tensor $Q^{\rho}{}_{\mu\nu}$.
Furthermore, in a relativistic framework, the mass of a test particle remains constant when it is defined with respect to the proper time.
Therefore, it may be difficult to construct a consistent proper time while keeping the mass constant in a spacetime with non-metricity.
In such a situation, the notion of particle mass should be regarded as an effective quantity rather than a strictly conserved invariant, and thus, it is natural to generalize the conventional test-particle action in STEGR.

For this reason, we adopt a generalized parameterization of the particle
action and introduce the following form:
\begin{align}
    I[x^{\mu}(\lambda)\,,\mathscr{E}(\lambda)] 
    = 
    \int^{\lambda_{2}}_{\lambda_{1}}d\lambda\left[\frac{1}{\mathscr{E}}\,g_{\mu\nu}\frac{dx^{\mu}}{d\lambda}\frac{dx^{\nu}}{d\lambda} + \sigma\,\mathscr{E}\right]
    \,,
\label{action for EoM}
\end{align}
where $\sigma$ is taken to be $-1$, $0$, and $+1$ for timelike, null, and spacelike curves, respectively, and $\mathscr{E}$ is a parameterization factor with $\mathscr{E} \neq 0\,$. 
This action is related to the standard proper-time particle action used in GR, but remains well-defined even when the norm of the four-momentum is not conserved. 
Related approaches have also been discussed in the literature.
Ref.~\cite{Iosifidis:2023eom} derives a possible test-particle motion based on the energy-momentum conservation.
In the present work, however, we restrict ourselves to the simplest setup.

Varying Eq.~\eqref{action for EoM} with respect to $\mathscr{E}$, we obtain
\begin{align}
    \sigma\,\mathscr{E}^{2} = g_{\mu\nu}\frac{dx^{\mu}}{d\lambda}\frac{dx^{\nu}}{d\lambda}\,.
\label{definition of length}
\end{align} 
The above formula defines the local distance between two different points, which leads to the Lagrangian for geodesic curve in GR with $\sigma = -1\,$: $\mathscr{E} = \sqrt{-g_{\mu\nu}(x)(\partial x^{\mu}/\partial \lambda)(\partial x^{\nu}/\partial \lambda)}$. 
In GR, we can always choose the Lorentz frame such that the Levi-Civita connection vanishes and \textcolor{blue}{$\mathscr{E}$} is a unity. 
In the timelike case, we obtain local Lorentz invariance, as desired by the Einstein equivalence principle. 
In general, a parameter $\lambda$ that satisfies $\dot{\mathscr{E}} = 0$ is distinguished as an affine parameter from others.\footnote{
Calculating $\dot{\mathscr{E}} = d\mathscr{E}/d\lambda = 0$ straightforwardly, we can derive the geodesic equation.
}
However, as we will show, we must modify the above descriptions in STEGR theories, except for the cases of $\sigma = 0\,$. 
Varying Eq.~\eqref{action for EoM} with respect to the coordinate functions $x^{\mu}$, we obtain 
\begin{align}
    \frac{d^{2}x^{\alpha}}{d\lambda^{2}} + \overset{\circ}{\Gamma}{}^{\alpha}_{\mu\nu}\frac{dx^{\mu}}{d\lambda}\frac{dx^{\nu}}{d\lambda} = \frac{1}{\mathscr{E}}\,\frac{de}{d\lambda}\,\frac{dx^{\alpha}}{d\lambda}\,.
\label{EoM of a scalar test particle}
\end{align}
In GR, we can take the parameter $\lambda$ as an affine parameter, which yields the desired geodesic equation. 

Differentiating Eq.~\eqref{definition of length} with respect to the parameter $\lambda$, we obtain
\begin{align}
    2\,\sigma\,\mathscr{E}\,\frac{d\mathscr{E}}{d\lambda} 
    = 
    \frac{dx^{\rho}}{d\lambda}\frac{dx^{\mu}}{d\lambda}\frac{dx^{\nu}}{d\lambda}Q_{\rho\mu\nu} + 2 g_{\mu\nu} \frac{dx^{\mu}}{d\lambda} \frac{dx^{\rho}}{d\lambda}\nabla_{\rho}\left(\frac{dx^{\nu}}{d\lambda}\right)
    \,,
\label{redefinition of parameter}
\end{align}
In a general metric-affine geometry, it is important to distinguish
between three notions: geodesics, defined as extremal curves of
the metric length, autoparallels defined by the affine connection
through $u^{\mu}\nabla_{\mu}u^{\nu}=0$, and the physical trajectory
of a particle.
In Riemannian geometry, these notions coincide because the connection is metric compatible.
However, in non-Riemannian geometries, they are generally distinct.

In \textit{non}-Riemannian geometries, the three concepts -- autoparallel, geodesic, and trajectory --  are independent as far as the particle has no coupling terms with geometric quantities. 
This independence raises a debatable issue regarding which type of curve a particle should follow in non-Riemannian theories. 
Let us consider two possible cases; 
i) we assume $du^{\nu}/d\lambda=u^{\mu}\nabla_{\mu}u^{\nu}=0\,$ is satisfied; 
ii) we do not assume $du^{\nu}/d\lambda=u^{\mu}\nabla_{\mu}u^{\nu}=0\,$, where $u^{\mu} \underset{\rm set}{:=} dx^{\mu}/d\lambda\,$.

First, we consider case i) as follows. 
In this case, for $\sigma = -1, 1$, we no longer choose an affine parameter. 
Combining Eq.~\eqref{EoM of a scalar test particle} with Eq.~\eqref{redefinition of parameter}, we obtain the equation of motion of the test scalar particle:
\begin{align}
    \frac{d^{2}x^{\alpha}}{d\lambda^{2}} 
    + \overset{\circ}{\Gamma}{}^{\alpha}_{\mu\nu}\frac{dx^{\mu}}{d\lambda}\frac{dx^{\nu}}{d\lambda} 
    = 
    q\,\frac{dx^{\alpha}}{d\lambda}
    \,,
\label{EoM of test scalar particle in STEGR theories in generic frame}
\end{align}
where $q$ is defined as
\begin{align}
    q 
    \underset{\rm set}{:=} 
    \frac{1}{2}\,\frac{u^{\rho}u^{\mu}u^{\nu}Q_{\rho\mu\nu}}{g_{\mu\nu}u^{\mu}u^{\nu}}
    \,.
\label{definition of q}
\end{align}
$Q_{\rho\mu\nu}$ is the non-metricity tensor, $Q_{\rho\mu\nu} \underset{\rm set}{\:=} \nabla_{\rho}g_{\mu\nu}\,$. 
$q$ is nothing more than the (half-)rate of change in the \mbox{(squared-)length} of the velocity vector $u^{\mu}$ along the trajectory of the test scalar particle. 
Of course, if we consider the internal STEGR, $Q_{\rho\mu\nu}$ is given by the St\"{u}ckelberg fields as in Eq.~\eqref{non-metricity in internal STEGR in Formalism 3}. 
In CGR, we employ the non-metricity in the form of Eq.~\eqref{non-metricity in coincident gauge}. 
In particular, Eq.~\eqref{mimetic CG in Palatini approach} is applicable at least in a global region in which no coordinate singularity exists, since the spacetime is flat and torsion-free in the affine-geometric sense.

For $\sigma = -1\,,1$, we conclude the following equation:
\begin{align}
    \frac{du^{\alpha}}{d\lambda} = q\,u^{\alpha}
    \, ,
\label{EoM of test scalar particle in STEGR theories}
\end{align}
and a test scalar particle obeys \textit{the equation of norm-flow} in STEGR theories.\footnote{
Eq.~\eqref{EoM of test scalar particle in STEGR theories in generic frame} is named after the fact that the non-metricity indicates a norm of the vector changes along the parallel transport. 
}
In the Weyl geometry, a special case of flat and torsion-free geometries, a consistent result is demonstrated in Refs.~\cite{Wada2023, Wada:2025szz} based on information geometry~\cite{Amari2016}. 
We remark that the above result does not depend on the signature $\sigma$.
Moreover, we do not fix the parameterization factor $\mathscr{E}$ in the derivation, suggesting that the change in the local distance is allowed under Eq.~\eqref{definition of length}. 
In other words, a parametrization based on a conserved norm, such as the proper time, is no longer available in STEGR theories.

Second, we consider case ii). 
In this case, we can take an affine parameter. 
That is, based on Eq.~(\ref{redefinition of parameter}), we can impose 
\begin{align}
     u^{\rho}\nabla_{\rho}u^{\sigma} = -\frac{1}{2}g^{\sigma\mu}u^{\rho}u^{\nu}Q_{\rho\mu\nu} \,.
\label{EoM of test scalar particle in STEGR theories in generic frame with affne parameter}
\end{align}
Under this condition, Eq.~(\ref{EoM of a scalar test particle}) becomes
\begin{align}
    \frac{d^{2}x^{\alpha}}{d\lambda^{2}} = 0\,.
\label{geodesic equation in STEGR theories}
\end{align}
Eq.~\eqref{geodesic equation in STEGR theories} is none other than the geodesic equation in the flat and torsion-free spacetime. 
However, this equation does not provide the trajectory of the test scalar particle.
The possible equation for the trajectory should be the modified autoparallel equation Eq.~\eqref{EoM of test scalar particle in STEGR theories in generic frame with affne parameter}.
This equation differs from the generalized autoparallel equation proposed in the pioneering works~\cite{Adak:2008gd, Adak:2011ltj}.

In the case of $\sigma = 0\,$, Eq.~\eqref{definition of length} suggests that $\dot{\mathscr{E}} = d\mathscr{E}/d\lambda$ takes arbitrary values. 
Thus, in this case, we can choose the parameter as an affine parameter: $\dot{\mathscr{E}} = 0\,$ and obtain
\begin{align}
    \frac{du^{\alpha}}{d\lambda} = 0\,.
\label{EoM of test scalar particle in STEGR theories in null sigma case}
\end{align}
Eq.~\eqref{EoM of test scalar particle in STEGR theories in null sigma case} is actually the geodesic equation in a flat and torsion-free spacetime. 
In other words, the length of the velocity vector of a massless test scalar particle does not change in parallel transportation, suggesting no difference from the result in GR. 

Our conclusion is different from that in the pioneering works on STEGR theories~\cite{Adak:2008gd, Adak:2011ltj}.
From our perspective, the equation in Refs.~\cite{Adak:2008gd, Adak:2011ltj} is appropriate for describing the trajectory of a test scalar particle in TEGR theories. 
If the spacetime is flat and non-metricity free, $q = 0$ by virtue of Eq.~\eqref{definition of q}. 
Thus, the parameter $\lambda$ becomes an affine parameter thanks to Eq.~\eqref{redefinition of parameter}, and we obtain 
\begin{align}
    \frac{d^{2}x^{\alpha}}{d\lambda^{2}} + \overset{\circ}{\Gamma}{}^{\alpha}_{\mu\nu}\frac{dx^{\mu}}{d\lambda}\frac{dx^{\nu}}{d\lambda} = 0\,.
\label{geodesic equation}
\end{align}
In the flat and non-metricity-free spacetime, there exists a frame in which $\overset{\circ}{\Gamma}{}^{\alpha}_{\mu\nu}$ vanishes, leading us to a case similar to STEGR.
In this case, we obtain 
\begin{align}
    \frac{d^{2}x^{\alpha}}{d\lambda^{2}} = 0\,.
\label{geodesic equation2}
\end{align}
This equation simply reflects the flat affine structure of the spacetime, in which the geodesic equation reduces to a trivial form.
However, this equation does not necessarily determine the physical
trajectory of the test particle in the presence of non-metricity.
At first sight, this result may appear puzzling since the equation takes the form of a straight-line solution.
To clarify this point, we recall that in a general affine spacetime, the parallel transport of a vector may depend on the path connecting two points.
This geometric property prevents us from the above seemingly contradictory result.

This strange result is ascribed to the independence between the two notions in a metric-affine differential geometry: the minimalness in the length between two points (`geodesic' curves) and the invariance of the direction of a vector (`autoparallel' curves). 
These two notions are the same as each other in (pseudo-)Riemannian geometry. 
In TEGR and STEGR theories, however, we should distinguish these two notions when considering the kinematics of a test scalar particle. 
These observations motivate us to consider the autoparallel equation as a natural candidate for describing the trajectory of a test scalar particle in TEGR theories,
\begin{align}
     u^{\mu}\nabla_{\mu}u^{\alpha} = \frac{d^{2}x^{\alpha}}{d\tau^{2}} + \Gamma^{\alpha}_{\mu\nu}\frac{dx^{\mu}}{d\tau}\frac{dx^{\nu}}{d\tau} = 0
     \,.
\label{EoM of TEGR}
\end{align}
$\tau$ is an affine parameter, which is guaranteed by the absence of non-metricity. 
Thus, it is a natural consideration to generalize the autoparallel equation to describe the motion of a test scalar particle in TEGR theories. 
This conclusion is also compatible with the result derived in Ref.~\cite{Capozziello:2022zzh}.\footnote{
Our results differ from the results provided in Ref.~\cite{Iosifidis:2023eom} in the following two points; 
1) Inclusion of coupling terms between a test particle and geometric quantities; 
2) Usage of the parameter for describing the trajectory of a test particle. 
Regarding the first point, we can fix this disagreement by introducing appropriate coupling terms into Eq.~\eqref{action for EoM} in our work. 
However, the second point distinguishes our work from Ref.~\cite{Iosifidis:2023eom} since the proper time is implicitly employed to derive the equation of motion, building on Refs.~\cite{Weinberg:1972kfs, Papapetrou:1974gq, Adler:1965}.
}
For the affine connection $\Gamma^{\alpha}_{\mu\nu}$ in Eq.~\eqref{EoM of TEGR}, we can choose the Weitzenb\"{o}ck connection given in Eq.~\eqref{Weitzenboch connection}. 

\section{\label{05}Conclusions}

In this work, we have revisited CGR based on the internal STEGR in Formalism 3, which was recently proposed by one of the authors in Ref.~\cite{Tomonari:2024vij}. 
We reviewed the two approaches to the STEGR theories: the Palatini and gauge approaches. 
In the Palatini approach, we identify the geometry by imposing two conditions, teleparallel and torsion-free conditions, via Lagrange multipliers. 
On the other hand, in the gauge approach, we realize these restrictions by fixing the gauge. 
However, regarding STEGR, there was no comprehensive understanding of how to establish this theory based on the gauge approach. 

We considered this issue and reformulated it correctly in the gauge approach. 
The key point to reconcile this issue was the generalization of the internal metric, as suggested in Refs.~\cite{Hu:2023gui, Tomonari:2024vij}. 
Then, we derived the field equations of the internal STEGR in Formalism 3 and clarified that the (Ostrogradski) ghost DOFs may exist. 
Thus, we must identify a degenerate condition by performing the DB analysis.
After that, we revisited CGR based on our formulation. 
Imposing the coincident gauge, we derived CGR in both terms of the action integral and the field equation. 

We also discussed the possible kinematics of the STEGR theories and derived a concrete equation of motion for a test scalar particle: the norm-flow equation Eq.~\eqref{EoM of test scalar particle in STEGR theories} and the modified autoparallel equation Eq.~\eqref{EoM of test scalar particle in STEGR theories in generic frame with affne parameter} for massive test scalar particles and the geodesic Eq.~\eqref{geodesic equation2} for massless test scalar particles. 
We emphasize that our formulation naturally incorporates the possibility of a variable mass arising from non-metricity, implying the existence of a coordinate system where the mass remains constant even without employing the proper time.
A thorough investigation of this issue is left for future work.
We summarize our results in Table~\ref{table: Summarize of revisiting CGR}.
\begin{table}[ht!]
    \centering
    \renewcommand{\arraystretch}{1.5}\caption{
    We summarize our work in a table. CGR is reformulated based on the internal space with the coincident gauge (`C-gauged internal space'). 
    As long as we restrict our consideration to a global region in spacetime such that a regular coordinate system exists, C-gauged internal space coincides with the global region in spacetime $\mathcal{M}\,$. 
    In particular, if we take the global region as spacetime itself, the formulation turns out to be that on spacetime (see Sec.~\ref{03:02} for details). 
    In STEGR theories, a massive test scalar particle obeys the norm-flow. 
    A massless test scalar particle respects the geodesics. 
    In detail, see Sec.~\ref{04}. In the internal STEGR, we have partially revealed that the spacetime metric plays a role as a multiplier. However, we should clarify the unknown bi-metric structure as a remaining issue.
    }
    \begin{tabular}{ c || c | c | c | c}
        Theory & Formulation & Configuration variable & Well-posed formulation & Equation of motion \\ \hline\hline
        GR & Spacetime & $g_{\mu\nu}$ & $\checkmark$ & geodesic \\ \hline
        Covariant TEGR & Internal-space & $e^{I}{}_{\mu}\,,\omega^{I}{}_{J\mu}$ & $\checkmark$ & autoparallel \\ \hline 
        TEGR & Internal-space & $e^{I}{}_{\mu}$ & $\checkmark$ & autoparallel \\ \hline
        Covariant STEGR & Spacetime & $g_{\mu\nu}$ and $\Gamma^{\rho}_{\mu\nu}$ & $\checkmark$ & 
        \begin{tabular}{c}
        massless: geodesic \\ / others: norm-flow or \\
        modified autoparalell
        \end{tabular}
        \\ \hline
        CGR & \textit{C-gauged Internal-space} & $g_{\mu\nu}$ & $\checkmark$ (Achieved in Sec.~\ref{03}) & 
        \begin{tabular}{c}
        massless: geodesic \\ /others: norm-flow or \\
        modified autoparalell
        \end{tabular}
        \\ \hline
        Internal STEGR & Internal-space & 
        $g_{IJ}$ and $\xi^{I}$ & $\triangle$ (See Refs.~\cite{Hu:2023gui, Tomonari:2024vij}) & 
        \begin{tabular}{c}
        massless: geodesic \\ /others: norm-flow or \\
        modified autoparalell
        \end{tabular}
        \\ \hline
    \end{tabular}
    \label{table: Summarize of revisiting CGR}
\end{table}

There was considerable confusion about the use of the coincident gauge. 
That is, the gauge-fixing method was applied misleadingly to the Palatini approach of STEGR when deriving CGR. 
Needless to say, one can fix the gauge in the gauge approach only. 
The crucial points to reconcile this issue are that the violation (or possibly extension) of the local Lorentz symmetry breaks the first-order formalism of GR. 
This violation/extension allows us to consider the generalization of the internal metric to a more generic one, which differs from the Minkowski metric~\cite{Hu:2023gui, Tomonari:2023ars, Tomonari:2024vij}. 
In this regard, the necessity of reconsidering the causal structure, the understanding of the light cone, is also suggested in Ref.~\cite{Obukhov:2024evf}. 
Their perspective is consistent with our conclusion. 
Based on the internal STEGR in Formalism 3, this work reconciles the confusion by providing a correct construction as a gauge approach to gravity.  

For future investigations, we should first elucidate the bi-metric structure of the internal STEGR in Formalism 3. 
Although our theory was enough to derive CGR in the gauge approach method, the configuration space should be spanned only by the gauge fields. 
In Sec.~\ref{03:01}, we mentioned that the spacetime metric behaves as if a Lagrange multiplier. 
If this is true, the bi-metric structure is simply an extension of the configuration space by an outer constraint, and it is always possible; however, further investigation is mandatory. 
To clarify this point, we should perform the DB analysis to reveal the constraint structure; that is, the symmetry of the theory, possible degenerate conditions to remove ghost modes, and the DOFs of the theory. 
Our theory is manifestly diffeomorphism-invariant, suggesting that the issue indicated in Ref.~\cite{DAmbrosio:2023asf} is remedied by taking the ADM foliation properly~\cite{Tomonari:2023wcs}.

Regarding the DOF analysis, the DB analysis of STEGR in both the spacetime and internal-space formulations has not yet been completed.
The DB analysis was performed in CGR, as was applied to the nonlinear extension of CGR, Coincident $f(Q)$ gravity~\cite{DAmbrosio:2020nqu, Hu:2022anq, Tomonari:2023wcs, Heisenberg:2023lru}. 
However, Ref.~\cite{DAmbrosio:2023asf} suggested an issue with the DB analysis in application to Teleparallel Theories of Gravity (TTG). 
Spatial boundary terms in the total Hamiltonian of a given TTG theory and Poisson brackets cause unsolvable partial differential equations (PDEs) of multipliers, and such PDEs break the Dirac algorithm down. 
Fortunately, if the theory is diffeomorphism invariant, the problematic spatial boundary terms vanish with the proper ADM foliation, allowing us to remedy the Dirac algorithm~\cite{Tomonari:2023wcs}. 
If this is not the case, bifurcation occurs due to broken symmetry, and the prescription can remedy only a specific sector of the system. 
Thus, it is mandatory to establish alternative methods for unveiling the constraint structure of a theory with the broken diffeomorphism invariance. 
In CGR, we inevitably encounter this issue~\cite{DAmbrosio:2023asf, Tomonari:2023wcs} since the diffeomorphism invariance is violated. 
The current work may contribute to the DOF analysis based on the internal-space formulation and gauge approach.

Moreover, it is mandatory to revisit the cosmological applications.
Refs.~\cite{Gomes:2023tur, Heisenberg:2023wgk} revealed that 
$f(Q)$ gravity theories suffer from a crucial pathology in the cosmological application. 
The theory exhibits a propagating ghost mode on the non-trivial branch I
and encounters the strong coupling issue on the trivial branch and non-trivial branch II in its linear perturbation~\cite{Gomes:2023hyk, Gomes:2023tur, Heisenberg:2023wgk}.
However, existing works concluded that these pathological features in $f(Q)$ gravity theories arise from the symmetry of $ISO(3)$ and its subgroup. 
Thus, if there exists a gauge-fixing condition that breaks the assumed symmetry and remedies these pathologies, $f(Q)$ gravity theories can be revived.
We can refine the argument about symmetry and gauge-fixing in $f(Q)$ gravity by applying the results of this work to the STEGR.
As another possibility, if a viable screening mechanism~\cite{Brax:2013ida, Brax:2021wcv} exists, $f(Q)$ gravity theories may also revive as a physics theory.
Connecting to the DOF analysis, we need to further investigate the fundamental DOF and propagation modes.

Finally, we will discuss the potential applications of our results to physical measurements. 
The norm-flow or modified autoparallel equation shows explicit deviation from an ordinary geodesic equation for the test scalar particle, and the non-metricity of the spacetime can be interpreted as an external force in Eq.~\eqref{EoM of test scalar particle in STEGR theories in generic frame}.
Taking the Newtonian limit, we can derive the modification of Newton's second law in a non-relativistic environment and examine it within the framework of well-known Solar System constraints on modified gravity theories.
Eq.~\eqref{EoM of test scalar particle in STEGR theories in generic frame} indicates that the modification due to the non-metricity is of the higher order of the velocity, and we expect that the observation does not strongly exclude the non-metricity.
However, it is still necessary to examine the Solar System constraint quantitatively.
We leave them for future work.

\acknowledgments

The authors thank Sousuke Noda, Friedrich W. Hehl, Takeru Asaka, and anonymous reviewers for giving beneficial comments on this work. 
K.T. thanks the Interfaculty Initiative in Information Studies, Graduate School of Interdisciplinary Information Studies, The University of Tokyo, for supporting this work.
T.K. is supported by the National Science Foundation of China (No.~12403003), and the National Key R\&D Program of China (No.~2021YFA0718500).

\appendix

\section{Explicit formulae of coefficients in \texorpdfstring{Eq.~\eqref{Field eqs of internal STEGR in formalism 3}}{TEXT} }
\label{app:01}

The entities of Eq.~\eqref{Field eqs of internal STEGR in formalism 3} are given explicitly as follows:
\begin{align}
\begin{split}
    &F^{(0)\,A}{}_{B}{}^{|\alpha\beta} := -\xi\,g^{\nu\alpha}\,g_{BC}\,\left(g^{\mu\rho}g^{DC}{}_{\mu\nu}\partial_{\rho}\xi_{D}\,\partial^{\beta}\xi^{A} + g^{\mu\beta}\tilde{g}^{AC}{}_{\mu\nu}\right)\,,\\
    &F^{(1)} := -\xi\,g^{\mu\alpha}g^{\nu\beta}\,g^{AB}{}_{\mu\nu}\,\partial_{\alpha}\xi_{A}\,\partial_{\beta}\xi_{B} \,,\\
    &F^{(2)\,A}{}_{BC}{}^{|\mu\nu\rho} := -\xi\,g^{\nu\alpha}\,g^{\rho\beta}\,\partial^{\mu}\xi^{A}\,\partial_{\alpha}\xi_{B}\,\partial_{\beta}\xi_{C} \,,\\
    &F^{(3)}{}_{A}{}^{|\mu\nu\rho} := -\xi\,g^{\mu\nu}g^{\rho\alpha}\,\partial_{\alpha}\xi_{A} \,,\\
    &F^{(4)\,A} := -\partial_{\beta}\xi_{C}\,\left\{\tilde{g}^{AC}{}_{\mu\nu}\,\partial_{\rho}\left(\xi\,g^{\mu\rho}\,g^{\nu\beta}\right) + g^{BC}{}_{\mu\nu}\,\partial_{\alpha}\xi_{B}\,\partial^{\rho}\xi^{A}\,\partial_{\rho}\left(\xi\,g^{\mu\alpha}\,g^{\nu\beta}\right)\right\}\\
    &\quad\quad\quad\quad\quad\quad - F^{(0)\,A}{}_{D}{}^{|\alpha\beta}\,g^{DB}\,\partial_{\alpha}g_{BC}\,\partial_{\beta}\xi^{C} + F^{(1)}\,\partial_{\alpha}g^{\alpha\beta}\,\partial_{\beta}\xi^{A} \,,
\end{split}
\label{entities of Field eqs wrt xi of internal STEGR formalism 3}
\end{align}
and
\begin{align}
\begin{split}
    &G^{(0)\,CDEF|\mu\nu} := g^{\mu\alpha}g^{\nu\beta}\,\partial_{\alpha}\xi_{A}\,\partial_{\beta}\xi_{B}\,\Phi^{ABCDEF}\,,\\
    &G^{(1)\,ABC|\mu\nu} := \Phi^{DCABEF}\,g^{\mu\alpha}g^{\nu\beta}\,\partial_{\alpha}\xi_{D}\,\partial_{\beta}g_{EF}\,,\\
    &G^{(2)\,ABC} := \Phi^{CDABEF}\,g^{\mu\nu}\,\partial_{\mu}\xi_{D}\,\partial_{\nu}g_{EF} \,,\\
    &G^{(3)\,AB} := \partial_{\alpha}\xi_{C}\,\partial_{\beta}\xi_{D}\,\partial_{\mu}\left(\Phi^{CDABEF}\,g^{\mu\alpha}g^{\nu\beta}\right)\,\partial_{\nu}g_{EF} + G^{(2)\,ABC}\,\partial_{\alpha}g^{\alpha\beta}\,\partial_{\beta}\xi_{C}\\ 
    &\quad - \xi\,g^{\mu\alpha}g^{\nu\beta}\left\{\frac{1}{4}\,g^{AB}g^{IJ}{}_{\mu\nu}\,\partial_{\alpha}\xi_{I}\,\partial_{\beta}\xi_{J} - \sqrt{-\mathrm{det}{(g_{AB}})}\,g^{AI}g^{BJ}g^{CD}g^{EF}\left(\Psi_{FCIEJD|\beta\alpha\mu\nu} + \tilde{\Psi}_{FICEJD|\beta\alpha\mu\nu}\right)\right\} \,.
\end{split}
\label{entities of Field eqs wrt eta of internal STEGR formalism 3}
\end{align}
Here, we define $\Psi_{ABCDEF|\alpha\beta\mu\nu}$, $\tilde{\Psi}_{ABCDEF|\alpha\beta\mu\nu}$, and $\Phi^{ABCDEF}$ as follows:
\begin{align}
\begin{split}
    &\quad \Psi_{ABCDEF|\alpha\beta\mu\nu} := g_{CDEF|\mu\nu}\,\partial_{\alpha}\xi_{A}\,\partial_{\beta}\xi_{B} - \frac{1}{2}\,g_{BFDA|\mu\nu}\,\partial_{\beta}\xi_{C}\,\partial_{\alpha}\xi_{E}\,,\\
    &\quad \tilde{\Psi}_{ABCDEF|\alpha\beta\mu\nu} := \tilde{g}_{CDEF|\mu\nu}\,\partial_{\alpha}\xi_{A}\,\partial_{\beta}\xi_{B} - \frac{1}{2}\,\tilde{g}_{BAFD|\mu\nu}\,\partial_{\beta}\xi_{C}\,\partial_{\alpha}\xi_{E}\,,\\
    &\quad \Phi^{ABCDEF} := -\frac{1}{2}\,\xi\,\sqrt{-\mathrm{det}(g_{AB})}\left\{-g^{ABCDEF} + \frac{1}{2}\left(g^{ABCEDF} + g^{ABFDEC}\right)\right\}\,.
\end{split}
\label{where of entities of Field eqs wrt eta of internal STEGR formalism 3}
\end{align}

\section{Explicit formulae of coefficients in \texorpdfstring{Eq.~\eqref{Field eqs of CGR without BTs}}{TEXT} }
\label{app:02}

The entities of Eq.~\eqref{Field eqs of CGR without BTs} are given explicitly as follows:
\begin{align}
\begin{split}
    &H^{\alpha\beta} := -\frac{1}{2}\,\partial_{\alpha}\left(\sqrt{-g}h^{\alpha\beta\mu\nu\lambda\kappa}\right)\overset{\rm c}{Q}_{\beta\lambda\kappa} - \sqrt{-g}\,g^{\alpha\beta\mu\nu\lambda\kappa}\,g^{\mu\nu}\,\overset{\rm c}{Q}_{\alpha\rho[\sigma|}\overset{\rm c}{Q}_{\beta|\lambda]\kappa} + 2\,\sqrt{-g}\,g^{\lambda\kappa}g^{\rho\sigma}g^{\mu\alpha}g^{\nu\beta}Q_{\mu\nu\rho\sigma\lambda\kappa} \,\\
    &\quad\quad\quad = -\frac{1}{2}\,\partial_{\alpha}\left(\sqrt{-g}h^{\alpha\beta\mu\nu\lambda\kappa}\right)\partial_{\beta}g_{\lambda\kappa} - \sqrt{-g}\,g^{\alpha\beta\mu\nu\lambda\kappa}\,g^{\mu\nu}\,\partial_{\alpha}g_{\rho[\sigma|}\partial_{\beta}g_{|\lambda]\kappa} + 2\,\sqrt{-g}\,g^{\lambda\kappa}g^{\rho\sigma}g^{\mu\alpha}g^{\nu\beta}\,Q_{\mu\nu\rho\sigma\lambda\kappa}\,,
\end{split}
\label{entities of Field eqs of CGR without BTs}
\end{align}
where we set $h^{\rho\sigma\mu\nu\lambda\kappa}$, $g^{\rho\sigma\mu\nu\lambda\kappa}$, and $Q_{\mu\nu\rho\sigma\lambda\kappa}$ as follows:
\begin{align}
\begin{split}
    &\quad h^{\rho\sigma\mu\nu\lambda\kappa} := -\frac{1}{2}\,\sqrt{-g}\,\left\{-g^{\rho\sigma\mu\nu\lambda\kappa} + \frac{1}{2}\left(g^{\rho\sigma\mu\lambda\nu\kappa} + g^{\rho\sigma\nu\lambda\kappa\mu}\right)\right\}\,,\\
    &\quad g^{\rho\sigma\mu\nu\lambda\kappa} := g^{\rho\sigma}g^{\mu\nu}g^{\lambda\kappa} + 2g^{\rho\kappa}g^{\sigma\nu}g^{\mu\lambda} \,,\\
    &\quad Q_{\mu\nu\rho\sigma\lambda\kappa} := \overset{\rm c}{Q}_{\mu\rho[\sigma|}\overset{\rm c}{Q}_{\nu|\lambda]\kappa} + 2\,\overset{\rm c}{Q}_{\rho\mu[\kappa|}\overset{\rm c}{Q}_{\lambda|\nu]\sigma} + 2\,\overset{\rm c}{Q}_{\lambda\rho[\sigma|}\overset{\rm c}{Q}_{\kappa|\mu]\nu} + 4\,\overset{\rm c}{Q}_{\mu\rho[\lambda|}\overset{\rm c}{Q}_{\kappa|\sigma]\nu}\,\\
    &\quad\quad\quad\quad\quad = \partial_{\mu}g_{\rho[\sigma|}\partial_{\nu}g_{|\lambda]\kappa} + 2\,\partial_{\rho}g_{\mu[\kappa|}\partial_{\lambda}g_{|\nu]\sigma} + 2\,\partial_{\lambda}g_{\rho[\sigma|}\partial_{\kappa}g_{|\mu]\nu} + 4\,\partial_{\mu}g_{\rho[\lambda|}\partial_{\kappa}g_{|\sigma]\nu}\,.
\end{split}
\label{where of entities of Field eqs of CGR without BTs}
\end{align}

\section{Explicit formulae of coefficients in \texorpdfstring{Eq.~\eqref{Field eqs of internal STEGR in formalism 3}}{TEXT} with the coincident gauge}
\label{app:03}

In the coincident gauge, Eq.~\eqref{entities of Field eqs wrt xi of internal STEGR formalism 3} turns out to be as follows:
\begin{align}
\begin{split}
    &F^{(0)\,A}{}_{B}{}^{|\alpha\beta} \underset{\rm C-gauge}{=} -\sqrt{-g}\,\delta^{A}{}_{\lambda}\,\delta_{B}{}^{\kappa}\,g^{\alpha\nu}\,g_{\kappa\rho}\,\left(g^{\beta\lambda}\,g^{\mu\rho\chi_{1}\chi_{2}\chi_{3}\chi_{4}}\,g_{\chi_{1}\chi_{2}\chi_{3}\chi_{4}|\mu\nu} + g^{\mu\beta}\,g^{\lambda\rho\chi_{1}\chi_{2}\chi_{3}\chi_{4}}\,\tilde{g}_{\chi_{1}\chi_{2}\chi_{3}\chi_{4}|\mu\nu}\right) \,,\\
    &F^{(1)\,} \underset{\rm C-gauge}{=} -\sqrt{-g}\,g^{\mu\nu\chi_{1}\chi_{2}\chi_{3}\chi_{4}}\,g_{\chi_{1}\chi_{2}\chi_{3}\chi_{4}|\mu\nu} \,,\\
    &F^{(2)\,A}{}_{BC}{}^{|\mu\nu\rho} \underset{\rm C-gauge}{:=} -\delta^{A}{}_{\lambda}\,\delta_{B}{}^{\nu}\,\delta_{C}{}^{\rho}\,g^{\mu\lambda} \,,\\
    &F^{(3)\,}{}_{A}{}^{|\mu\nu\rho} \underset{\rm C-gauge}{:=} -\delta_{A}{}^{\rho}\,g^{\mu\nu} \,,\\
    &F^{(4)\,A} \underset{\rm C-gauge}{:=} -\sqrt{-g}\,\delta^{A}{}_{\lambda}\,\left[\tilde{g}_{\chi_{1}\chi_{2}\chi_{3}\chi_{4}|\mu\nu}\,\left\{g_{\beta\gamma}\,g^{\lambda\gamma\chi_{1}\chi_{2}\chi_{3}\chi_{4}}\,\partial_{\rho}\left(g^{\mu\rho}g^{\nu\beta}\right) -g^{\alpha\nu}g^{\beta\mu}g^{\lambda\rho\chi_{1}\chi_{2}\chi_{3}\chi_{4}}\,\partial_{\alpha}g_{\beta\rho}\right\} \right.\\
    &\quad \left. + g_{\chi_{1}\chi_{2}\chi_{3}\chi_{4}|\mu\nu}\left\{g_{\beta\gamma}g_{\alpha\kappa}g^{\rho\lambda}\,g^{\kappa\gamma\chi_{1}\chi_{2}\chi_{3}\chi_{4}}\,\partial_{\rho}\left(g^{\mu\alpha}g^{\nu\beta}\right) - g^{\alpha\nu}g^{\beta\lambda}g^{\mu\rho\chi_{1}\chi_{2}\chi_{3}\chi_{4}}\,\partial_{\alpha}g_{\beta\rho} + g^{\mu\nu\chi_{1}\chi_{2}\chi_{3}\chi_{4}}\,\partial_{\alpha}g^{\alpha\lambda}\right\}\right] \,,
\end{split}
\label{}
\end{align}
where we set $g_{\alpha\beta\rho\lambda|\mu\nu}$ and $\tilde{g}_{\alpha\beta\rho\lambda|\mu\nu}$ as follows:
\begin{align}
\begin{split}
    &\quad g_{\alpha\beta\rho\lambda|\mu\nu} := \overset{\rm c}{Q}_{\mu\alpha[\beta|}\overset{\rm c}{Q}_{\nu|\rho]\lambda} = \partial_{\mu}g_{\alpha[\beta|}\partial_{\nu}g_{|\rho]\lambda} \,,\\
    &\quad \tilde{g}_{\alpha\beta\rho\lambda|\mu\nu} := \overset{\rm c}{Q}_{\mu\alpha\beta}\overset{\rm c}{Q}_{\nu\rho\lambda} - \overset{\rm c}{Q}_{(\mu|\alpha\rho}\overset{\rm c}{Q}_{|\nu)\beta\lambda} = \partial_{\mu}g_{\alpha\beta}\partial_{\nu}g_{\rho\lambda} - \partial_{(\mu|}g_{\alpha\rho}\partial_{|\nu)}g_{\beta\lambda}\,.
\end{split}
\label{}
\end{align}
Eq.~\eqref{entities of Field eqs wrt eta of internal STEGR formalism 3} turns out to be as follows:
\begin{align}
\begin{split}
    &G^{(0)\,CDEF|\mu\nu} \underset{\rm C-gauge}{=} \delta^{\mu}{}_{A}\,\delta^{\nu}{}_{B}\,\Phi^{ABCDEF} \,,\\
    &G^{(1)\,ABC|\mu\nu} \underset{\rm C-gauge}{=} \delta^{ABC}{}_{\alpha\beta\gamma}\,h^{\mu\gamma\alpha\beta\lambda\kappa}\,g^{\nu\rho}\,\overset{\rm c}{Q}_{\rho\lambda\kappa} = \delta^{ABC}{}_{\alpha\beta\gamma}\,h^{\mu\gamma\alpha\beta\lambda\kappa}\,g^{\nu\rho}\,\partial_{\rho}g_{\lambda\kappa} \,,\\
    &G^{(2)\,ABC} \underset{\rm C-gauge}{=} \delta^{ABC}{}_{\alpha\beta\gamma}\,h^{\alpha\beta\gamma\rho\lambda\kappa}\,\overset{\rm c}{Q}{}_{\rho\lambda\kappa} = \delta^{ABC}{}_{\alpha\beta\gamma}\,h^{\alpha\beta\gamma\kappa\rho\lambda}\,\partial_{\kappa}g_{\rho\lambda}\,,\\
    &G^{(3)\,AB} \underset{\rm C-gauge}{=} g_{\alpha\gamma}\,g_{\beta\rho}\,\partial_{\nu}g_{\lambda\kappa}\partial_{\mu}\left(\delta_{C}{}^{\gamma}\,\delta_{D}{}^{\rho}\,\delta_{E}{}^{\lambda}\,\delta_{F}{}^{\kappa}\,\Phi^{CDABEF}\,g^{\mu\alpha}\,g^{\nu\beta}\right) \\
    &\quad\quad\quad\quad\quad\quad + \delta_{C}{}^{\gamma}\,g^{\alpha\beta}\,\delta^{ABC}{}_{\chi_{1}\chi_{2}\chi_{3}}h^{\chi_{1}\chi_{2}\chi_{3}\kappa_{1}\kappa_{2}\kappa_{3}}\,\overset{\rm c}{Q}_{\kappa_{1}\kappa_{2}\kappa_{3}}\,\overset{\rm c}{Q}_{\alpha\beta\gamma}  \\
    &\quad\quad\quad\quad\quad\quad + \sqrt{-g}\,\delta^{A}{}_{\chi_{1}}\,\delta^{B}{}_{\chi_{2}}\,\Big{[}-\frac{1}{4}\,g^{\chi_{1}\chi_{2}}\,g^{\mu\nu\chi_{1}\chi_{2}\chi_{3}\chi_{4}}\,g_{\chi_{1}\chi_{2}\chi_{3}\chi_{4}|\mu\nu}\\ 
    &\quad\quad\quad\quad\quad\quad\quad\quad\quad\quad\quad\quad\quad\quad + g^{\chi_{1}\chi_{5}}\,g^{\chi_{2}\chi_{6}}\,g^{\chi_{3}\chi_{7}}\,g^{\chi_{4}\chi_{8}}\,\left(\tilde{g}_{\chi_{3}\chi_{4}\chi_{6}\chi_{7}|\chi_{8}\chi_{5}} - \frac{1}{2}\,\tilde{g}_{\chi_{5}\chi_{8}\chi_{7}\chi_{4}|\chi_{6}\chi_{3}}\right)\Big{]} \,.
\end{split}
\label{}
\end{align}
Here, we define $\Psi_{ABCDEF|\alpha\beta\mu\nu}$, $\tilde{\Psi}_{ABCDEF|\alpha\beta\mu\nu}$, $\Phi^{ABCDEF}$, $\delta^{ABCDEF}{}_{\alpha\beta\rho\lambda\gamma\delta}$, and $\delta^{ABC}{}_{\mu\nu\rho}$ as follows:
\begin{align}
\begin{split}
    &\quad \Psi_{ABCDEF|\alpha\beta\mu\nu} \underset{\rm C-gauge}{=}\\
    &\quad\quad \left(\delta_{C}{}^{\chi_{1}}\,\delta_{D}{}^{\chi_{2}}\,\delta_{E}{}^{\chi_{3}}\,\delta_{F}{}^{\chi_{4}}\,\delta_{A}{}^{\kappa_{1}}\,\delta_{B}{}^{\kappa_{2}} - \frac{1}{2}\,\delta_{B}{}^{\chi_{1}}\,\delta_{F}{}^{\chi_{2}}\,\delta_{D}{}^{\chi_{3}}\,\delta_{A}{}^{\chi_{4}}\,\delta_{E}{}^{\kappa_{1}}\,\delta_{C}{}^{\kappa_{2}}\right)\,g_{\alpha\kappa_{1}}\,g_{\beta\kappa_{2}}\,g_{\chi_{1}\chi_{2}\chi_{3}\chi_{4}|\mu\nu} \,,\\
    &\quad \tilde{\Psi}_{ABCDEF|\alpha\beta\mu\nu} \underset{\rm C-gauge}{=}\\
    &\quad\quad \left(\delta_{C}{}^{\chi_{1}}\,\delta_{D}{}^{\chi_{2}}\,\delta_{E}{}^{\chi_{3}}\,\delta_{F}{}^{\chi_{4}}\,\delta_{A}{}^{\kappa_{2}}\,\delta_{B}{}^{\kappa_{1}} - \frac{1}{2}\,\delta_{B}{}^{\chi_{1}}\,\delta_{A}{}^{\chi_{2}}\,\delta_{F}{}^{\chi_{3}}\,\delta_{D}{}^{\chi_{4}}\,\delta_{E}{}^{\kappa_{2}}\,\delta_{C}{}^{\kappa_{1}}\right)\,g_{\alpha\kappa_{1}}\,g_{\beta\kappa_{2}}\,\tilde{g}_{\chi_{1}\chi_{2}\chi_{3}\chi_{4}|\mu\nu} \,,\\
    &\quad \Phi^{ABCDEF} \underset{\rm C-gauge}{=} -\frac{1}{2}\,\sqrt{-g}\left[-\delta^{ABCDEF}{}_{\alpha\beta\rho\lambda\gamma\delta} + \frac{1}{2}\left(\delta^{ABCEDF}{}_{\alpha\beta\rho\lambda\gamma\delta} + \delta^{ABFDEC}{}_{\alpha\beta\rho\lambda\gamma\delta}\right)\right]\,g^{\alpha\beta}g^{\rho\lambda}g^{\gamma\delta} \,,\\
    &\quad \delta^{ABCDEF}{}_{\alpha\beta\rho\lambda\gamma\delta} := \delta^{A}{}_{\alpha}\,\delta^{B}{}_{\beta}\,\delta^{C}{}_{\rho}\,\delta^{D}{}_{\lambda}\,\delta^{E}{}_{\gamma}\,\delta^{F}{}_{\delta} + 2\,\delta^{A}{}_{\alpha}\,\delta^{F}{}_{\beta}\,\delta^{B}{}_{\rho}\,\delta^{D}{}_{\lambda}\,\delta^{C}{}_{\gamma}\,\delta^{E}{}_{\delta} \,,\\
    &\quad \delta^{ABC}{}_{\mu\nu\rho} := \delta^{A}{}_{\mu}\,\delta^{B}{}_{\nu}\,\delta^{C}{}_{\rho} \,.
\end{split}
\label{}
\end{align}

\bibliography{bibliography_revisitingCGR.bib}

\providecommand{\href}[2]{#2}\begingroup\raggedright\begin{thebibliography}{10}

\bibitem{Nester:1998mp}
J.~M. Nester and H.-J. Yo, ``{Symmetric teleparallel general relativity},'' {\em Chin. J. Phys.} {\bf 37} (1999)  113, \href{http://arxiv.org/abs/gr-qc/9809049}{{\tt arXiv:gr-qc/9809049}}.

\bibitem{Bahamonde:2021gfp}
S.~Bahamonde, K.~F. Dialektopoulos, C.~Escamilla-Rivera, G.~Farrugia, V.~Gakis, M.~Hendry, M.~Hohmann, J.~Levi~Said, J.~Mifsud, and E.~Di~Valentino, ``{Teleparallel gravity: from theory to cosmology},'' \href{http://dx.doi.org/10.1088/1361-6633/ac9cef}{{\em Rept. Prog. Phys.} {\bf 86} (2023) no.~2, 026901}, \href{http://arxiv.org/abs/2106.13793}{{\tt arXiv:2106.13793 [gr-qc]}}.

\bibitem{Hehl:1994ue}
F.~W. Hehl, J.~D. McCrea, E.~W. Mielke, and Y.~Ne'eman, ``{Metric affine gauge theory of gravity: Field equations, Noether identities, world spinors, and breaking of dilation invariance},'' \href{http://dx.doi.org/10.1016/0370-1573(94)00111-F}{{\em Phys. Rept.} {\bf 258} (1995)  1--171}, \href{http://arxiv.org/abs/gr-qc/9402012}{{\tt arXiv:gr-qc/9402012}}.

\bibitem{Kiefer:2004xyv}
C.~Kiefer, {\em {Quantum gravity}}, vol.~124.
\newblock Clarendon, Oxford, 2004.

\bibitem{Hawking1973}
S.~W. Hawking and G.~F.~R. Ellis, {\em The Large Scale Structure of Space-Time}.
\newblock Cambridge Monographs on Mathematical Physics. Cambridge University Press, 1973.

\bibitem{Wald:1984rg}
R.~M. Wald, \href{http://dx.doi.org/10.7208/chicago/9780226870373.001.0001}{{\em {General Relativity}}}.
\newblock Chicago University Press, 1984.

\bibitem{BeltranJimenez:2017tkd}
J.~Beltr\'an~Jim\'enez, L.~Heisenberg, and T.~Koivisto, ``{Coincident General Relativity},'' \href{http://dx.doi.org/10.1103/PhysRevD.98.044048}{{\em Phys. Rev. D} {\bf 98} (2018) no.~4, 044048}, \href{http://arxiv.org/abs/1710.03116}{{\tt arXiv:1710.03116 [gr-qc]}}.

\bibitem{Heisenberg:2023lru}
L.~Heisenberg, ``{Review on f(Q) gravity},'' \href{http://dx.doi.org/10.1016/j.physrep.2024.02.001}{{\em Phys. Rept.} {\bf 1066} (2024)  1--78}, \href{http://arxiv.org/abs/2309.15958}{{\tt arXiv:2309.15958 [gr-qc]}}.

\bibitem{Gomes:2023hyk}
D.~A. Gomes, J.~Beltr\'an~Jim\'enez, and T.~S. Koivisto, ``{General parallel cosmology},'' \href{http://dx.doi.org/10.1088/1475-7516/2023/12/010}{{\em JCAP} {\bf 12} (2023)  010}, \href{http://arxiv.org/abs/2309.08554}{{\tt arXiv:2309.08554 [gr-qc]}}.

\bibitem{Gomes:2023tur}
D.~A. Gomes, J.~Beltr\'an~Jim\'enez, A.~J. Cano, and T.~S. Koivisto, ``{Pathological Character of Modifications to Coincident General Relativity: Cosmological Strong Coupling and Ghosts in f(Q) Theories},'' \href{http://dx.doi.org/10.1103/PhysRevLett.132.141401}{{\em Phys. Rev. Lett.} {\bf 132} (2024) no.~14, 141401}, \href{http://arxiv.org/abs/2311.04201}{{\tt arXiv:2311.04201 [gr-qc]}}.

\bibitem{Heisenberg:2023wgk}
L.~Heisenberg, M.~Hohmann, and S.~Kuhn, ``{Cosmological teleparallel perturbations},'' \href{http://dx.doi.org/10.1088/1475-7516/2024/03/063}{{\em JCAP} {\bf 03} (2024)  063}, \href{http://arxiv.org/abs/2311.05495}{{\tt arXiv:2311.05495 [gr-qc]}}.

\bibitem{Weitzenboh1923}
R.~Weitzenboch, ``Invarianten theorie,'' {\em Nordhoff, Groningen} (1923)  320.

\bibitem{Obukhov:2002tm}
Y.~N. Obukhov and J.~G. Pereira, ``{Metric affine approach to teleparallel gravity},'' \href{http://dx.doi.org/10.1103/PhysRevD.67.044016}{{\em Phys. Rev. D} {\bf 67} (2003)  044016}, \href{http://arxiv.org/abs/gr-qc/0212080}{{\tt arXiv:gr-qc/0212080}}.

\bibitem{Ferraro:2016wht}
R.~Ferraro and M.~J. Guzm\'an, ``{Hamiltonian formulation of teleparallel gravity},'' \href{http://dx.doi.org/10.1103/PhysRevD.94.104045}{{\em Phys. Rev. D} {\bf 94} (2016) no.~10, 104045}, \href{http://arxiv.org/abs/1609.06766}{{\tt arXiv:1609.06766 [gr-qc]}}.

\bibitem{Blagojevic:2023fys}
M.~Blagojevi\'c and J.~M. Nester, ``{From the Lorentz invariant to the coframe form of f(T) gravity},'' \href{http://dx.doi.org/10.1103/PhysRevD.109.064034}{{\em Phys. Rev. D} {\bf 109} (2024) no.~6, 064034}, \href{http://arxiv.org/abs/2312.14603}{{\tt arXiv:2312.14603 [gr-qc]}}.

\bibitem{Einstein1928}
A.~Einstein, ``Riemann-geometrie mit aufrechterhaltung des begriffes des fernparallelismus,'' {\em Preussische Akademie der Wissenschaften, Phys.Math. Klasse, Sitzungsberichte.} (1928)  217.

\bibitem{Tomonari:2024vij}
K.~Tomonari, ``{STEGR in internal-space formulation: Formalisms, primary constraints, and possible internal symmetries},'' \href{http://dx.doi.org/10.1063/5.0253087}{{\em J. Math. Phys.} {\bf 66} (2025) no.~5, 052505}, \href{http://arxiv.org/abs/2410.04848}{{\tt arXiv:2410.04848 [gr-qc]}}.

\bibitem{Tomonari:2023ars}
K.~Tomonari, ``{A unified-description of curvature, torsion, and non-metricity of the metric-affine geometry with the M\"obius representation},'' \href{http://dx.doi.org/10.1142/S021988782450333X}{{\em Int. J. Geom. Meth. Mod. Phys.} {\bf 22} (2025) no.~05, 2450333}, \href{http://arxiv.org/abs/2312.11558}{{\tt arXiv:2312.11558 [gr-qc]}}.

\bibitem{Blagojevic:2000pi}
M.~Blagojevic and M.~Vasilic, ``{Gauge symmetries of the teleparallel theory of gravity},'' \href{http://dx.doi.org/10.1088/0264-9381/17/18/313}{{\em Class. Quant. Grav.} {\bf 17} (2000)  3785--3798}, \href{http://arxiv.org/abs/hep-th/0006080}{{\tt arXiv:hep-th/0006080}}.

\bibitem{Capozziello:2022zzh}
S.~Capozziello, V.~De~Falco, and C.~Ferrara, ``{Comparing equivalent gravities: common features and differences},'' \href{http://dx.doi.org/10.1140/epjc/s10052-022-10823-x}{{\em Eur. Phys. J. C} {\bf 82} (2022) no.~10, 865}, \href{http://arxiv.org/abs/2208.03011}{{\tt arXiv:2208.03011 [gr-qc]}}.

\bibitem{Stueckelberg:1938a}
J.~Lacki, H.~Ruegg, and G.~Wanders, eds., {\em Die Wechselwirkungs Kr{\"a}fte in der Elektrodynamik und in der Feldtheorie der Kernkraefte (I) [39]}, \href{http://dx.doi.org/10.1007/978-3-7643-8878-2_16}{pp.~251--271}.
\newblock Birkh{\"a}user Basel, Basel, 2009.
\newblock \url{https://doi.org/10.1007/978-3-7643-8878-2_16}.

\bibitem{Stueckelberg:1938b}
J.~Lacki, H.~Ruegg, and G.~Wanders, eds., {\em Die Wechselwirkungskr{\"a}fte in der Elektrodynamik und in der Feldtheorie der Kernkr{\"a}fte (Teil II und III) [40]}, \href{http://dx.doi.org/10.1007/978-3-7643-8878-2_17}{pp.~273--303}.
\newblock Birkh{\"a}user Basel, Basel, 2009.
\newblock \url{https://doi.org/10.1007/978-3-7643-8878-2_17}.

\bibitem{Ruegg:2003ps}
H.~Ruegg and M.~Ruiz-Altaba, ``{The Stueckelberg field},'' \href{http://dx.doi.org/10.1142/S0217751X04019755}{{\em Int. J. Mod. Phys. A} {\bf 19} (2004)  3265--3348}, \href{http://arxiv.org/abs/hep-th/0304245}{{\tt arXiv:hep-th/0304245}}.

\bibitem{Adak:2008gd}
M.~Adak, O.~Sert, M.~Kalay, and M.~Sari, ``{Symmetric Teleparallel Gravity: Some exact solutions and spinor couplings},'' \href{http://dx.doi.org/10.1142/S0217751X13501674}{{\em Int. J. Mod. Phys. A} {\bf 28} (2013)  1350167}, \href{http://arxiv.org/abs/0810.2388}{{\tt arXiv:0810.2388 [gr-qc]}}.

\bibitem{Adak:2011ltj}
M.~Adak and C.~Pala, ``{A novel approach to autoparallels for the theories of symmetric teleparallel gravity},'' \href{http://dx.doi.org/10.1088/1742-6596/2191/1/012017}{{\em J. Phys. Conf. Ser.} {\bf 2191} (2022) no.~1, 012017}, \href{http://arxiv.org/abs/1102.1878}{{\tt arXiv:1102.1878 [physics.gen-ph]}}.

\bibitem{DAmbrosio:2020nqu}
F.~D'Ambrosio, M.~Garg, L.~Heisenberg, and S.~Zentarra, ``{ADM formulation and Hamiltonian analysis of Coincident General Relativity},'' \href{http://arxiv.org/abs/2007.03261}{{\tt arXiv:2007.03261 [gr-qc]}}.

\bibitem{BeltranJimenez:2022azb}
J.~Beltr\'an~Jim\'enez and T.~S. Koivisto, ``{Lost in translation: The Abelian affine connection (in the coincident gauge)},'' \href{http://dx.doi.org/10.1142/S0219887822501080}{{\em Int. J. Geom. Meth. Mod. Phys.} {\bf 19} (2022) no.~07, 2250108}, \href{http://arxiv.org/abs/2202.01701}{{\tt arXiv:2202.01701 [gr-qc]}}.

\bibitem{BeltranJimenez:2019esp}
J.~Beltr\'an~Jim\'enez, L.~Heisenberg, and T.~S. Koivisto, ``{The Geometrical Trinity of Gravity},'' \href{http://dx.doi.org/10.3390/universe5070173}{{\em Universe} {\bf 5} (2019) no.~7, 173}, \href{http://arxiv.org/abs/1903.06830}{{\tt arXiv:1903.06830 [hep-th]}}.

\bibitem{Motohashi:2016ftl}
H.~Motohashi, K.~Noui, T.~Suyama, M.~Yamaguchi, and D.~Langlois, ``{Healthy degenerate theories with higher derivatives},'' \href{http://dx.doi.org/10.1088/1475-7516/2016/07/033}{{\em JCAP} {\bf 07} (2016)  033}, \href{http://arxiv.org/abs/1603.09355}{{\tt arXiv:1603.09355 [hep-th]}}.

\bibitem{Olmo:2011uz}
G.~J. Olmo, ``{Palatini Approach to Modified Gravity: f(R) Theories and Beyond},'' \href{http://dx.doi.org/10.1142/S0218271811018925}{{\em Int. J. Mod. Phys. D} {\bf 20} (2011)  413--462}, \href{http://arxiv.org/abs/1101.3864}{{\tt arXiv:1101.3864 [gr-qc]}}.

\bibitem{Dirac:1950pj}
P.~A.~M. Dirac, ``{Generalized Hamiltonian dynamics},'' \href{http://dx.doi.org/10.4153/CJM-1950-012-1}{{\em Can. J. Math.} {\bf 2} (1950)  129--148}.

\bibitem{Dirac:1958sc}
P.~A.~M. Dirac, ``{The Theory of gravitation in Hamiltonian form},'' \href{http://dx.doi.org/10.1098/rspa.1958.0142}{{\em Proc. Roy. Soc. Lond. A} {\bf 246} (1958)  333--343}.

\bibitem{Anderson:1951ta}
J.~L. Anderson and P.~G. Bergmann, ``{Constraints in covariant field theories},'' \href{http://dx.doi.org/10.1103/PhysRev.83.1018}{{\em Phys. Rev.} {\bf 83} (1951)  1018--1025}.

\bibitem{Bergmann:1949zz}
P.~G. Bergmann, ``{Non-Linear Field Theories},'' \href{http://dx.doi.org/10.1103/PhysRev.75.680}{{\em Phys. Rev.} {\bf 75} (1949)  680--685}.

\bibitem{Bergmann1950}
P.~G. Bergmann, R.~Penfield, R.~Schiller, and H.~Zatzkis, ``The {H}amiltonian of the general theory of relativity with electromagnetic field,'' \href{http://dx.doi.org/10.1103/PhysRev.80.81}{{\em Phys.Rev.} {\bf 80} (1950)  81}.

\bibitem{BergmannBrunings1949}
P.~G. Bergmann and J.~H.~M. Brunings, ``Non-linear field theories {II}. {C}anonical equations and quantization,'' \href{http://dx.doi.org/10.1103/RevModPhys.21.480}{{\em Rev.Mod.Phys.} {\bf 21} (1949)  480}.

\bibitem{Sugano:1982bm}
R.~Sugano and T.~Kimura, ``{On the Relation of First Class Constraints to Gauge Degrees of Freedom},'' \href{http://dx.doi.org/10.1143/PTP.69.252}{{\em Prog. Theor. Phys.} {\bf 69} (1983)  252}.

\bibitem{Sugano:1989rq}
R.~Sugano and T.~Kimura, ``{Gauge Transformations for Dynamical Systems With First and Second Class Constraints},'' \href{http://dx.doi.org/10.1103/PhysRevD.41.1247}{{\em Phys. Rev. D} {\bf 41} (1990)  1247}.

\bibitem{Sugano:1991ir}
R.~Sugano, Y.~Kagraoka, and T.~Kimura, ``{On gauge transformations and gauge fixing conditions in constraint systems},'' \href{http://dx.doi.org/10.1142/S0217751X92000041}{{\em Int. J. Mod. Phys. A} {\bf 7} (1992)  61--90}.

\bibitem{Ong:2017xwo}
Y.~C. Ong and J.~M. Nester, ``{Counting Components in the Lagrange Multiplier Formulation of Teleparallel Theories},'' \href{http://dx.doi.org/10.1140/epjc/s10052-018-6050-3}{{\em Eur. Phys. J. C} {\bf 78} (2018) no.~7, 568}, \href{http://arxiv.org/abs/1709.00068}{{\tt arXiv:1709.00068 [gr-qc]}}.

\bibitem{JimenezCano:2021rlu}
A.~Jim\'enez~Cano, {\em {Metric-affine Gauge theories of gravity. Foundations and new insights}}.
\newblock PhD thesis, Granada U., Theor. Phys. Astrophys., 2021.
\newblock \href{http://arxiv.org/abs/2201.12847}{{\tt arXiv:2201.12847 [gr-qc]}}.

\bibitem{Hu:2023gui}
K.~Hu, M.~Yamakoshi, T.~Katsuragawa, S.~Nojiri, and T.~Qiu, ``{Nonpropagating ghost in covariant f(Q) gravity},'' \href{http://dx.doi.org/10.1103/PhysRevD.108.124030}{{\em Phys. Rev. D} {\bf 108} (2023) no.~12, 124030}, \href{http://arxiv.org/abs/2310.15507}{{\tt arXiv:2310.15507 [gr-qc]}}.

\bibitem{Rosen:1940zza}
N.~Rosen, ``{General Relativity and Flat Space. I},'' \href{http://dx.doi.org/10.1103/PhysRev.57.147}{{\em Phys. Rev.} {\bf 57} (1940)  147--150}.

\bibitem{Rosen:1940zz}
N.~Rosen, ``{General Relativity and Flat Space. II},'' \href{http://dx.doi.org/10.1103/PhysRev.57.150}{{\em Phys. Rev.} {\bf 57} (1940)  150--153}.

\bibitem{Hassan:2011zd}
S.~F. Hassan and R.~A. Rosen, ``{Bimetric Gravity from Ghost-free Massive Gravity},'' \href{http://dx.doi.org/10.1007/JHEP02(2012)126}{{\em JHEP} {\bf 02} (2012)  126}, \href{http://arxiv.org/abs/1109.3515}{{\tt arXiv:1109.3515 [hep-th]}}.

\bibitem{Schmidt-May:2015vnx}
A.~Schmidt-May and M.~von Strauss, ``{Recent developments in bimetric theory},'' \href{http://dx.doi.org/10.1088/1751-8113/49/18/183001}{{\em J. Phys. A} {\bf 49} (2016) no.~18, 183001}, \href{http://arxiv.org/abs/1512.00021}{{\tt arXiv:1512.00021 [hep-th]}}.

\bibitem{DAmbrosio:2023asf}
F.~D'Ambrosio, L.~Heisenberg, and S.~Zentarra, ``{Hamiltonian Analysis of f(Q)$f(\mathbb {Q})$ Gravity and the Failure of the Dirac\textendash{}Bergmann Algorithm for Teleparallel Theories of Gravity},'' \href{http://dx.doi.org/10.1002/prop.202300185}{{\em Fortsch. Phys.} {\bf 71} (2023) no.~12, 2300185}, \href{http://arxiv.org/abs/2308.02250}{{\tt arXiv:2308.02250 [gr-qc]}}.

\bibitem{Tomonari:2023wcs}
K.~Tomonari and S.~Bahamonde, ``{Dirac\textendash{}Bergmann analysis and degrees of freedom of coincident f(Q)-gravity},'' \href{http://dx.doi.org/10.1140/epjc/s10052-024-12677-x}{{\em Eur. Phys. J. C} {\bf 84} (2024) no.~4, 349}, \href{http://arxiv.org/abs/2308.06469}{{\tt arXiv:2308.06469 [gr-qc]}}. [Erratum: Eur.Phys.J.C 84, 508 (2024)].

\bibitem{Einstein1916}
A.~Einstein, ``Hamilton's principle and the general theory of relativity,'' {\em Sitzungsber.Preuss.Akad.Wiss.Berlin (Math.Phys)} (1916)  1111.

\bibitem{Runkla:2018xrv}
M.~R{\"u}nkla and O.~Vilson, ``{Family of scalar-nonmetricity theories of gravity},'' \href{http://dx.doi.org/10.1103/PhysRevD.98.084034}{{\em Phys. Rev. D} {\bf 98} (2018) no.~8, 084034}, \href{http://arxiv.org/abs/1805.12197}{{\tt arXiv:1805.12197 [gr-qc]}}.

\bibitem{Gibbons:1976ue}
G.~W. Gibbons and S.~W. Hawking, ``{Action Integrals and Partition Functions in Quantum Gravity},'' \href{http://dx.doi.org/10.1103/PhysRevD.15.2752}{{\em Phys. Rev. D} {\bf 15} (1977)  2752--2756}.

\bibitem{York:1972sj}
J.~W. York, Jr., ``{Role of conformal three geometry in the dynamics of gravitation},'' \href{http://dx.doi.org/10.1103/PhysRevLett.28.1082}{{\em Phys. Rev. Lett.} {\bf 28} (1972)  1082--1085}.

\bibitem{Erdmenger:2023hne}
J.~Erdmenger, B.~He\ss{}, R.~Meyer, and I.~Matthaiakakis, ``{Gibbons-Hawking-York boundary terms and the generalized geometrical trinity of gravity},'' \href{http://dx.doi.org/10.1103/PhysRevD.110.066002}{{\em Phys. Rev. D} {\bf 110} (2024) no.~6, 066002}, \href{http://arxiv.org/abs/2304.06752}{{\tt arXiv:2304.06752 [hep-th]}}.

\bibitem{Wada2023}
T.~Wada, ``Weyl geometric approach to the gradient-flow equations in information geometry,'' \href{http://dx.doi.org/10.7546/jgsp-66-2023-59-70}{{\em Journal of Geometry and Symmetry in Physics} {\bf 66} (2023)  59--70}, \href{http://arxiv.org/abs/2212.14706}{{\tt arXiv:2212.14706 [math-ph]}}.

\bibitem{Wada:2025szz}
T.~Wada and S.~Noda, ``{Weyl symmetry of the gradient-flow in information geometry},'' \href{http://arxiv.org/abs/2502.03866}{{\tt arXiv:2502.03866 [gr-qc]}}.

\bibitem{Iosifidis:2023eom}
D.~Iosifidis and F.~W. Hehl, ``{Motion of test particles in spacetimes with torsion and nonmetricity},'' \href{http://dx.doi.org/10.1016/j.physletb.2024.138498}{{\em Phys. Lett. B} {\bf 850} (2024)  138498}, \href{http://arxiv.org/abs/2310.15595}{{\tt arXiv:2310.15595 [gr-qc]}}.

\bibitem{Amari2016}
S.~Amari, \href{http://dx.doi.org/10.1007/978-4-431-55978-8}{{\em Information Geometry and Its Applications}}.
\newblock Applied Mathematical Sciences. Springer Japan, 2016.

\bibitem{Weinberg:1972kfs}
S.~Weinberg, {\em {Gravitation and Cosmology}: {Principles and Applications of the General Theory of Relativity}}.
\newblock John Wiley and Sons, New York, 1972.

\bibitem{Papapetrou:1974gq}
A.~Papapetrou, {\em {Lectures on General Relativity}}.
\newblock D. REIDEL PUBLISHING COMPANY, DORDRECHT-HOLLAND / BOSTON U.S.A., 1974.

\bibitem{Adler:1965}
R.~Adler, M.~Bazin, M.~Schiffer, and J.~E. Romain, ``Introduction to general relativity,'' \href{http://dx.doi.org/10.1063/1.3047725}{{\em Physics Today} {\bf 18 (9)} (1965) no.~68, }.

\bibitem{Obukhov:2024evf}
Y.~N. Obukhov and F.~W. Hehl, ``{Violating Lorentz invariance minimally by the emergence of nonmetricity? A Perspective},'' \href{http://dx.doi.org/10.1002/andp.202400217}{{\em Annalen der Physik} (2024)  2400217}, \href{http://arxiv.org/abs/2409.19411}{{\tt arXiv:2409.19411 [gr-qc]}}.

\bibitem{Hu:2022anq}
K.~Hu, T.~Katsuragawa, and T.~Qiu, ``{ADM formulation and Hamiltonian analysis of f(Q) gravity},'' \href{http://dx.doi.org/10.1103/PhysRevD.106.044025}{{\em Phys. Rev. D} {\bf 106} (2022) no.~4, 044025}, \href{http://arxiv.org/abs/2204.12826}{{\tt arXiv:2204.12826 [gr-qc]}}.

\bibitem{Brax:2013ida}
P.~Brax, ``{Screening mechanisms in modified gravity},'' \href{http://dx.doi.org/10.1088/0264-9381/30/21/214005}{{\em Class. Quant. Grav.} {\bf 30} (2013)  214005}.

\bibitem{Brax:2021wcv}
P.~Brax, S.~Casas, H.~Desmond, and B.~Elder, ``{Testing Screened Modified Gravity},'' \href{http://dx.doi.org/10.3390/universe8010011}{{\em Universe} {\bf 8} (2021) no.~1, 11}, \href{http://arxiv.org/abs/2201.10817}{{\tt arXiv:2201.10817 [gr-qc]}}.

\end{thebibliography}\endgroup
\bibliographystyle{utphys}
\end{document}